\documentclass[final,3p,times,number]{elsarticle}
\usepackage{amssymb}
\usepackage{lipsum}
\usepackage{amsmath}
\usepackage{xcolor}
\usepackage{booktabs}
\usepackage{multirow}
\usepackage{float} 

\usepackage[colorlinks=true,
            linkcolor=blue,
            citecolor=blue,
            urlcolor=blue]{hyperref}

\journal{Physics Open}

\begin{document}

\begin{frontmatter}



\title{A Physics-based and Geometry-aware Reduced-order Framework for Accelerated Part-Scale Thermomechanical Simulation of Additive Friction Stir Deposition Process}


\author{Apurba Sarker} %% Author name
%\author[label1]{Sourav Saha}

\author{Sourav Saha\corref{cor1}}
\ead{souravsaha@vt.edu}
\cortext[cor1]{Corresponding author}

\affiliation{organization={Kevin T. Crofton Department of Aerospace and Ocean Engineering, Virginia Polytechnic Institute and State University},%Department and Organization
            addressline={1600 Innovation Drive}, 
            city={Blacksburg},
            postcode={24061}, 
            state={VA},
            country={United States}}
            
\begin{abstract}
Additive friction stir deposition (AFSD) is attractive for naval and aerospace sustainment because it enables high-rate, solid-state near-net-shape fabrication and repair of large metallic components. However, full part-scale thermomechanical simulations remain computationally demanding because of the moving heat source, progressive material deposition, and evolving geometry. This work presents a physics-based and geometry-aware reduced-order framework for rapid prediction of temperature and residual stress fields in part-scale AFSD builds. The transient temperature field is represented using a tensor-product quadratic B-spline basis on a fixed Cartesian background grid, while arbitrary part geometries are embedded using a signed distance function (SDF) and a smooth material indicator. Residual stresses are recovered through an inherent strain formulation in which an offline thermo-elasto-plastic representative volume element (RVE) simulation supplies a calibrated eigenstrain field that is spatially scaled using the part-scale peak-temperature history. The thermal formulation is evaluated against published numerical and experimental AFSD data and two sample multilayer builds are simulated as demonstrations. For a four-layer thin-wall build, the thermal solution requires approximately 21.6~s compared to 806~s for the finite element model, while the complete thermomechanical framework requires approximately 86.4~s compared to 2,406~s.  For a larger 20-layer hollow cylindrical build, the reduced-order thermal simulations requires 0.148-0.152~h compared with 1.8~h for the finite element model, corresponding to thermal speed-ups of approximately 11.8-12.2$\times$. The complete thermomechanical framework requires 0.346~h compared to 5.4~h for the full finite element analysis, providing an overall speed-up of approximately 15.6$\times$. A thorough analysis of the predicted temperature and residual stress fields is performed to rationalize the prediction. The proposed framework enables rapid part-scale prediction of both thermal histories and final residual stress fields without repeated body-fitted meshing, full part-scale thermo-elasto-plastic analysis, or a massive offline dataset enabling superior model predictive control and process design in AFSD.
\end{abstract}




\end{frontmatter}

%\tableofcontents

%\linenumbers


\section{Introduction}
\label{introduction}

Additive friction stir deposition (AFSD) is a solid-state metal additive manufacturing process~\cite{yu2018nonbeam} that uses a rotating tool to frictionally heat and severely plastically deform a metal rod against a substrate rather than melting a feedstock. This solid-state mechanism avoids many of the defects that are common in fusion-based additive processes~\cite{yu2021deformation}. Because melting is not involved, problems related to cracking and high residual stresses are significantly reduced, and alloys with high crack susceptibility can be processed successfully~\cite{yakubov2024evolution}. Such alloys often cannot be reliably processed by laser- or arc-based fusion methods~\cite{martin2017printing}. The severe plastic deformation at temperatures below the melting point also produces fine, equiaxed, recrystallized grains, giving AFSD-deposited components mechanical behavior similar to forged material~\cite{wu2024critical,phillips2019microstructure}. Another defining advantage of AFSD is its high build rate, which makes the process particularly suitable for the rapid fabrication and repair of large-scale metallic components~\cite{yu2018nonbeam,yu2021deformation}. Because of these combined capabilities, AFSD is applied to large-scale near-net-shape fabrication and structural repair, applications that are difficult for conventional beam-based additive processes to match~\cite{griffiths2019perspective}.

Although AFSD offers clear advantages, the process is not equally straightforward for every material or every geometry. Bonding and consolidation are governed by the coupled effects of tool rotational speed, axial force, traverse speed, and material feed rate. Therefore, a suitable processing window must be established for each material system~\cite{chen2023effect,systematic_review2024}. Almost all defects in AFSD can be traced back to a poorly constructed process window~\cite{systematic_review2024}. Insufficient heat generation results in inadequate material flow, incomplete mixing, and poor interfacial bonding, while excessive heat input causes heat accumulation, geometric distortion, and dimensional inaccuracy~\cite{kumar2026state}. Geometric freedom is also more restricted than in powder-bed fusion. The current AFSD tooling produces wide deposition tracks on the order of centimeters, and the process requires special fixtures to build intricate shapes~\cite{yu2021deformation,mishra2022review}, so AFSD is better described as a near-net-shape process rather than a route to fine, complex geometry. These constraints make the transient temperature history central to the process, because the local thermal cycle governs material flow, bonding quality, microstructural evolution, and the buildup of residual stress~\cite{hu2024thermal,zhu2024bayesian}. Residual stress prediction is therefore essential at the part scale, because residual stress governs distortion, structural integrity, and post-process performance~\cite{yakubov2024evolution}. 

Part-scale thermomechanical simulation of AFSD remains challenging because the process combines a moving heat source, continuously evolving deposited geometry, highly localized severe plastic deformation near the tool, and rapid material deposition~\cite{hu2024thermal,systematic_review2024,mishra2022review}. Several modeling strategies have been developed to reduce the computational cost of thermomechanical simulations in metal additive manufacturing. Conventional body-fitted finite-element models resolve the moving heat source, progressive layer activation, steep temperature gradients, and resulting plastic deformation. However, their computational cost becomes prohibitive when the complete build volume and process history are represented at high spatial and temporal resolution~\cite{denlinger2017thermomechanical,chiumenti2017numerical}. Inherent strain methods were therefore developed as a faster alternative, in which the local inelastic deformation history was replaced by a calibrated equivalent strain field imposed in a single elastic mechanical solve~\cite{setien2019empirical,liang2018modified}. This approach was later extended and modified to improve its accuracy and its treatment of thin-walled and lattice structures~\cite{liang2019modified,liang2020homogenization,dong2021new,to2023modified}. Reduced-order and immersed-boundary methods were also introduced to lower the cost of thermal prediction, including proper generalized decomposition, GPU-accelerated multilevel solvers, and immersed thermomechanical solvers that avoided body-fitted meshing altogether~\cite{chiumenti2019pgd,ganeriwala2024gomelt,carraturo2020modeling,carraturo2021immersed}. Related multiphysics frameworks have coupled process-scale thermal and fluid-flow simulations with microstructure-evolution models to establish direct process--structure relationships in metal manufacturing~\cite{Kanak2025}. These methods have improved computational efficiency considerably, but part-scale simulation of real, complex parts remains a challenge. As deposition progresses, the computational geometry continuously evolves through the addition of new material. This growth also changes the thermal boundary conditions, as previously exposed surfaces become internal interfaces and new surfaces are created. Consequently, the transient temperature field must be repeatedly solved throughout the expanding three-dimensional part and substrate, resulting in substantial computational cost~\cite{chiumenti2017numerical,gouge2021hot,moran2018superposition}. For AFSD specifically, this challenge is further compounded by the high material-deposition rate and the strongly coupled heat generation, severe plastic deformation, material flow, and mechanical consolidation occurring directly beneath the rotating tool~\cite{gotawala2023material,kincaid2023finite,ruvo2026surface}.

Recent studies have begun to investigate data-driven and physics-informed models for predicting AFSD temperature evolution and material-flow behavior \cite{zhu2024bayesian,shi2024afsdnets,wu2026sequential,kumar2026state}. However, the broader development of such models remains constrained by the availability of sufficiently large and physically representative datasets \cite{zhu2024bayesian,liao2023hybrid,wu2026sequential}. Experimental data generation is costly and generally provides measurements only at a limited number of spatial locations, whereas high-fidelity thermo-elasto-plastic simulations can provide complete field information but become computationally prohibitive for part-scale geometries and long deposition histories \cite{zhu2024bayesian,liao2023hybrid,ruvo2026surface, carraturo2020modeling,denlinger2017thermomechanical,gouge2021hot}. Consequently, conventional full-order models are not well suited to the repeated evaluations required for design-space exploration, process optimization, uncertainty quantification, or large-scale training-data generation \cite{liao2023hybrid,gouge2021hot,moran2018superposition, wu2026sequential}. This computational limitation also restricts the use of high-fidelity simulation as an online forward model for control-oriented applications. Model-predictive control, for example, requires repeated forward predictions within a limited decision interval, motivating the use of reduced-order or computationally inexpensive thermal models in additive-manufacturing control frameworks \cite{liaomcpherson2022inlayer, vanblitterswijk2025realtime}. A computationally efficient physics-based model is therefore needed not only for direct thermomechanical prediction, but also as a scalable data generator for surrogate development and as a forward model for future optimization and control frameworks.

The present framework represents the transient temperature field using a tensor-product quadratic B-spline basis rather than a full body-fitted mesh. This separable, tensor-product construction was previously developed in \cite{batley2025khronos}, where a compact quadratic B-spline representation was shown to give resource-efficient predictions. The same separable-architecture idea was later generalized as a primitive for unified predictive and generative modeling~\cite{batley2026sna}, and a related multi-fidelity B-spline approach was also used to accelerate airfoil design optimization~\cite{sarker2026kernel}. In the present work, this separable B-spline representation is adapted to the thermal field generated during AFSD. More broadly, smooth basis functions and local refinement have been shown in isogeometric analysis to represent complex fields efficiently without repeated meshing~\cite{hughes2005isogeometric,hollig2003finite}. The part geometry is embedded into a fixed Cartesian background grid using a signed distance function, following the same fictitious-domain philosophy used in finite cell and immersed boundary methods for additive manufacturing~\cite{parvizian2007finite,duster2008finite,carraturo2020modeling}, so no repeated meshing or remeshing is required as the part is built up layer by layer. The predicted peak temperature history is then converted into a local thermal severity measure, which scales an RVE-calibrated inherent strain field. The residual stress problem is solved as a single elastic equilibrium problem with imposed eigenstrain, following the same logic used in modified inherent strain models for laser powder bed fusion~\cite{liang2019modified,to2023modified}, rather than as a full thermo-elasto-plastic simulation at every point. This coupling is designed to keep the dominant part-scale thermal and residual stress behavior of AFSD while reducing the cost of direct simulation, and it is validated here against the surface-based heat source benchmark of Ruvo et al.~\cite{ruvo2026surface}.

Reducing computational time while retaining the dominant part-scale thermomechanical behavior is the primary contribution of the present work. Existing studies have introduced several approaches for accelerating thermomechanical simulations of metal additive manufacturing, including adaptive discretization, layer aggregation, inherent strain formulations, reduced-order models, and immersed-boundary methods~\cite{gouge2021hot,liang2019modified,chiumenti2019pgd,carraturo2020modeling,carraturo2021immersed}. However, many of these approaches have been demonstrated or validated using relatively simple geometries, particularly thin-wall or single-track structures containing only a limited number of deposited layers~\cite{li2019octree,burkhardt2022thermomechanical,seitz2024inherent}. Although such configurations are valuable for numerical assessment and experimental validation, industrially relevant components generally involve more complex three-dimensional geometries and builds containing tens to hundreds of layers. The present study therefore considers two complementary test cases: a four-layer wall deposit and a 20-layer hollow cylindrical deposit. The four-layer wall is used for experimental and numerical validation against previously published AFSD results~\cite{ruvo2026surface}, whereas the 20-layer hollow cylinder extends the proposed framework to a more complex part-scale geometry. Together, these cases assess both the accuracy and computational efficiency of the framework and demonstrate its potential applicability to industrial-scale AFSD simulations.

The remainder of this paper is organized as follows. Section~\ref{sec:computational_methodology} introduces the proposed offline--online thermomechanical framework, including the geometry-embedded B-spline thermal formulation and the RVE-calibrated inherent strain approach. Section~\ref{sec:validation} evaluates the thermal predictions against published numerical and experimental AFSD results. Section~\ref{sec:problem_formulation} defines the geometries, process parameters, thermal boundary conditions, RVE calibration procedures, and mechanical models for the four-layer thin-wall and 20-layer hollow cylindrical builds. The results are then presented for the four-layer thin-wall build in Section~\ref{sec:thin_wall} and for the 20-layer hollow cylindrical build in Section~\ref{sec:cylinder_results}, including computational performance, transient thermal behavior, process-parameter effects, inherent strain calibration, and final residual stress predictions. Section~\ref{sec:conclusions} summarizes the principal findings, and Section~\ref{sec:limitations} discusses the current limitations and directions for future development. The nomenclature, complete mathematical formulation, convergence assessment, and temperature-dependent material properties are provided in the appendices.

The principal contributions of this work are summarized as follows:

\begin{itemize}
    \item \textbf{Order-of-magnitude computational speed-up without sacrificing accuracy.} The coupled thermal-ROM/inherent strain framework achieves thermal speed-ups of approximately 11.8--12.2$\times$ for the 20-layer hollow cylindrical build, and full thermomechanical speed-ups of approximately 27.8$\times$ and 15.2--15.6$\times$, respectively, relative to full finite-element thermomechanical analysis. This computational advantage is achieved without a large offline training dataset, unlike data-driven surrogate approaches~\cite{zhu2024bayesian,shi2024afsdnets,wu2026sequential}, and is retained as the deposition history lengthens and the geometry becomes more complex from a straight four-layer wall validated directly against published experimental and finite-element AFSD results~\cite{ruvo2026surface}, to a 20-layer hollow cylinder with a closed circumferential path.

    \item \textbf{A continuously thermally-scaled modified inherent strain formulation.} The original modified inherent strain method~\cite{liang2019modified,liang2020homogenization,dong2021new,to2023modified} applied a single, spatially uniform inherent strain to an entire built or homogenized region. The present work instead couples the offline RVE-calibrated inherent strain to the online, part-scale, spatially resolved peak-temperature history predicted by the thermal ROM, through the local thermal-severity index $\chi(\mathbf{x})$ (Eq.~\eqref{eq:main_inherent_strain_mapping}). This produces a spatially continuous inherent strain field that varies point-by-point with the actual predicted thermal loading, rather than a single homogenized value, without requiring a full part-scale thermo-elasto-plastic simulation.

    \item \textbf{A data-free, physics-based framework validated across two geometric complexities.} The coupled thermal-ROM/inherent strain framework is demonstrated on both a four-layer thin-wall benchmark, validated directly against published experimental and finite element AFSD results~\cite{ruvo2026surface}, and a 20-layer hollow cylindrical build with a closed circumferential path, including a cylindrical RVE calibration extending the inherent strain calibration procedure beyond the prismatic geometries typically considered in the literature. Both cases achieve order-of-magnitude computational speed-ups relative to full finite element thermomechanical analysis without requiring a large offline training dataset, unlike data-driven surrogate approaches~\cite{zhu2024bayesian,shi2024afsdnets,wu2026sequential}.
\end{itemize}

\section{Computational Methodology}
\label{sec:computational_methodology}

\begin{figure}[h!]
\centering
\includegraphics[width=0.9\linewidth]{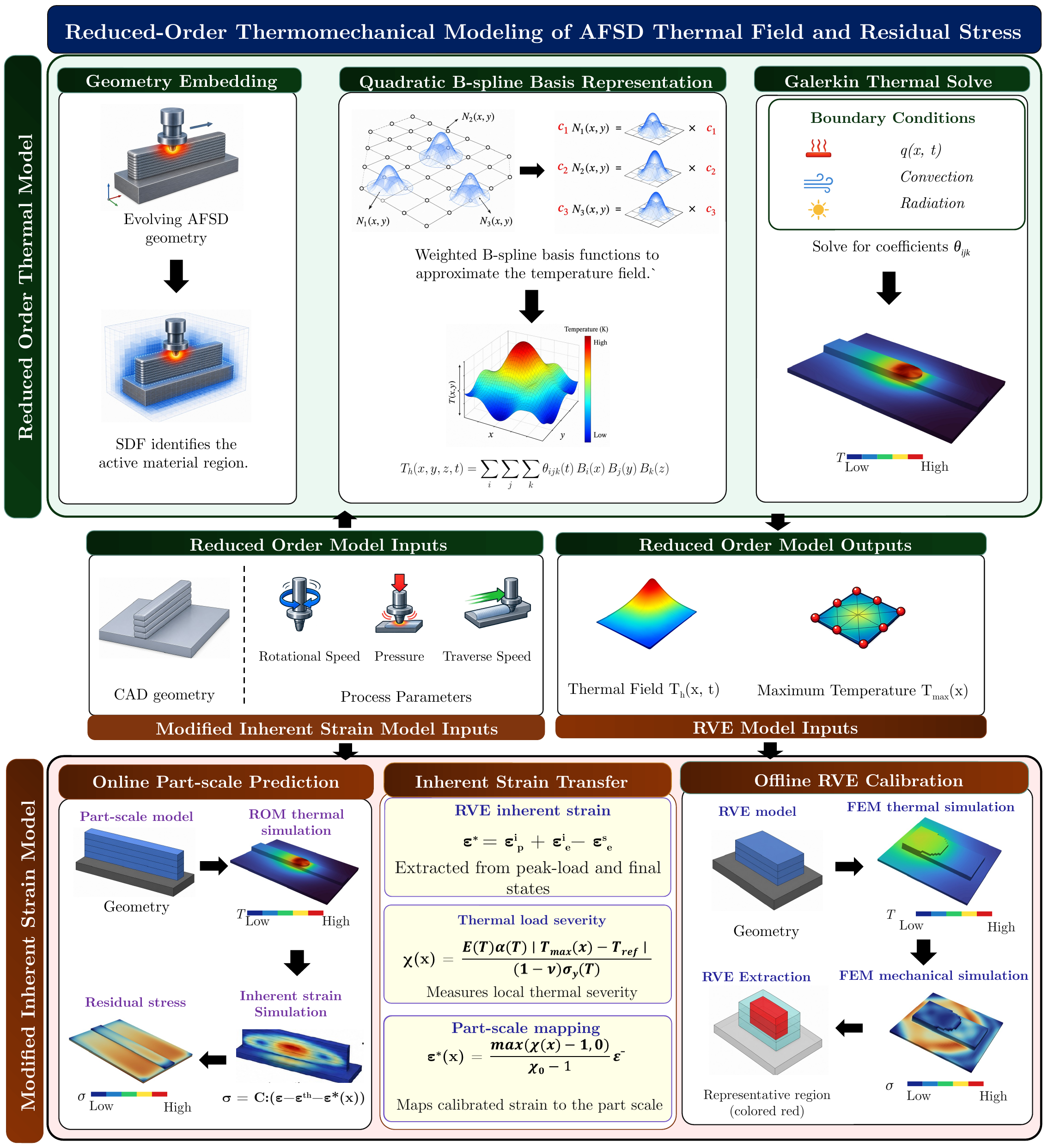}
\caption{Overview of the proposed thermomechanical framework. The left panel shows the geometry-embedded reduced-order thermal model, which predicts the transient temperature field on a fixed Cartesian grid using a quadratic B-spline Galerkin formulation. The right panel shows the modified inherent strain mechanical framework, where an offline thermo-elasto-plastic RVE simulation provides the representative inherent strain and the predicted thermal history is used to transfer it to the part scale for final residual stress prediction.}
\label{fig:overview}
\end{figure}

The proposed framework consists of two sequential components: a part-scale thermal simulation and a residual stress simulation, as illustrated in Figure~\ref{fig:overview}. The thermal simulation first predicts the transient temperature history throughout the evolving AFSD build. The resulting temperature field is then supplied to the residual stress model, which combines an offline representative-volume-element calibration with an online part-scale mechanical analysis.

\subsection{Geometry-Embedded B-Spline Thermal Formulation}
\label{subsec:thermal_formulation}

The transient temperature field during AFSD is governed by conservation of thermal energy over the time-dependent material domain $\Omega_m(t)$,

\begin{equation}
    \rho C_p(T)\frac{\partial T}{\partial t}
    =
    \nabla\cdot\left[k(T)\nabla T\right]
    +
    q(\mathbf{x},t),
    \qquad
    \mathbf{x}\in\Omega_m(t),
    \label{eq:main_heat_equation}
\end{equation}

where $\rho$ is the density, $C_p(T)$ is the temperature-dependent specific heat, $k(T)$ is the temperature-dependent thermal conductivity, and $q(\mathbf{x},t)$ represents the moving AFSD heat input. Heat loss from the exposed material boundary $\Gamma_m(t)$ is represented by convection and radiation,

\begin{equation}
    -k(T)\nabla T\cdot\mathbf{n}
    =
    h(T-T_{\infty})
    +
    \epsilon_{\mathrm{rad}}\sigma_{\mathrm{SB}}
    \left(T^4-T_{\infty}^4\right),
    \qquad
    \mathbf{x}\in\Gamma_m(t),
    \label{eq:main_thermal_boundary}
\end{equation}

where $\mathbf{n}$ is the outward unit normal, $h$ is the convective heat-transfer coefficient, $T_{\infty}$ is the ambient temperature, $\epsilon_{\mathrm{rad}}$ is the surface emissivity, and $\sigma_{\mathrm{SB}}$ is the Stefan--Boltzmann constant.

In a conventional body-fitted formulation, the computational mesh must conform to the evolving domains $\Omega_m(t)$ and $\Gamma_m(t)$. In the present formulation, the complete build is instead enclosed by a fixed Cartesian background domain $\Omega$, while the instantaneous material geometry is represented by a signed distance function $\phi(\mathbf{x},t)$. Negative values of $\phi$ identify material points, $\phi=0$ defines the material boundary, and positive values identify the exterior. Consequently, arbitrary or progressively deposited geometries can be represented by updating the scalar distance field without modifying the underlying Cartesian grid.

The signed distance field is converted into a smooth material indicator and a diffuse surface measure according to

\begin{equation}
    \eta(\mathbf{x},t)
    =
    H_{\epsilon}\!\left[-\phi(\mathbf{x},t)\right],
    \qquad
    \gamma(\mathbf{x},t)
    =
    \delta_{\epsilon}\!\left[\phi(\mathbf{x},t)\right]
    \left|\nabla\phi(\mathbf{x},t)\right|,
    \label{eq:main_geometry_embedding}
\end{equation}

where $\eta\approx1$ inside the active material and $\eta\approx0$ outside it, while $\gamma$ localizes surface contributions within a narrow region surrounding the embedded boundary. The regularized definitions of $H_{\epsilon}$ and $\delta_{\epsilon}$ are provided in~\ref{app:mathematical_formulation}.

Multiplying Eq.~\eqref{eq:main_heat_equation} by a test function $w$, applying integration by parts to the conduction term, and incorporating the boundary condition of Eq.~\eqref{eq:main_thermal_boundary} gives the weak thermal problem. Using $\eta$ to restrict material-volume contributions and $\gamma$ to represent the exposed material surface, the problem is evaluated entirely over the fixed Cartesian background domain as

\begin{align}
    \int_{\Omega}
    \rho C_p(T)\eta w
    \frac{\partial T}{\partial t}\,dV
    &+
    \int_{\Omega}
    k(T)\eta
    \nabla w\cdot\nabla T\,dV
    \nonumber\\
    &+
    \int_{\Omega}
    h\gamma w(T-T_{\infty})\,dV
    +
    \int_{\Omega}
    \epsilon_{\mathrm{rad}}\sigma_{\mathrm{SB}}
    \gamma w
    \left(T^4-T_{\infty}^4\right)\,dV
    \nonumber\\
    &=
    \int_{\Omega}
    \eta wq(\mathbf{x},t)\,dV .
    \label{eq:main_embedded_weak_form}
\end{align}

Equation~\eqref{eq:main_embedded_weak_form} provides the direct connection between the physical thermal problem and the embedded geometry representation. As deposition progresses, changes in the material volume and exposed boundary enter through the time-dependent fields $\eta(\mathbf{x},t)$ and $\gamma(\mathbf{x},t)$, while the computational grid itself remains unchanged.

The temperature field is approximated on this fixed domain using tensor-product quadratic B-spline basis functions,

\begin{equation}
    T_h(x,y,z,t)
    =
    \sum_{i=1}^{n_x}
    \sum_{j=1}^{n_y}
    \sum_{k=1}^{n_z}
    \theta_{ijk}(t)
    B_i(x)B_j(y)B_k(z),
    \label{eq:main_bspline_temperature}
\end{equation}

where $B_i(x)$, $B_j(y)$, and $B_k(z)$ are one-dimensional quadratic B-spline basis functions and $\theta_{ijk}(t)$ are the time-dependent temperature coefficients. Selecting the test functions from the same B-spline space and substituting Eq.~\eqref{eq:main_bspline_temperature} into Eq.~\eqref{eq:main_embedded_weak_form} produces the semi-discrete system

\begin{equation}
    \mathbf{M}(\boldsymbol{\theta})
    \dot{\boldsymbol{\theta}}
    +
    \mathbf{K}(\boldsymbol{\theta})
    \boldsymbol{\theta}
    +
    \mathbf{C}\boldsymbol{\theta}
    +
    \mathbf{r}_{\mathrm{rad}}(\boldsymbol{\theta})
    =
    \mathbf{f}(t),
    \label{eq:main_thermal_semidiscrete}
\end{equation}

where $\mathbf{M}$, $\mathbf{K}$, and $\mathbf{C}$ are the thermal-capacity, conduction, and convective heat-loss operators, respectively, $\mathbf{r}_{\mathrm{rad}}$ is the nonlinear radiation contribution, and $\mathbf{f}$ is the moving heat-input vector. The detailed operator definitions, Cartesian quadrature procedure, temporal discretization, and matrix-free implementation are provided in ~\ref{app:mathematical_formulation}.

\subsection{RVE-Calibrated Inherent Strain Formulation}

The second component of the framework predicts the final residual stress field using the modified inherent strain method, originally proposed by Liang et al.~\cite{liang2019modified} and further developed by~\cite{liang2020homogenization,dong2021new,to2023modified}. While the original formulation applied a spatially uniform inherent strain over the entire built region, the present work modifies this approach by scaling the RVE-calibrated inherent strain continuously in space according to the local peak-temperature history predicted by the part-scale thermal solver. This spatial scaling, introduced through the thermal-severity index $\chi(\mathbf{x})$ and the mapping of Eq.~\eqref{eq:main_inherent_strain_mapping}, allows regions that experienced different thermal loading during deposition to receive a correspondingly different inherent strain, rather than a single spatially invariant value. This component contains an offline RVE calibration stage and an online part-scale mechanical analysis. During the offline stage, a detailed thermo-elasto-plastic finite element simulation is performed on a representative volume element subjected to the local AFSD thermal and mechanical loading conditions. The representative inherent strain is extracted from the RVE response as
\begin{equation}
    \boldsymbol{\varepsilon}^{*}_{\mathrm{avg}}
    =
    \left\langle
    \boldsymbol{\varepsilon}^{i}_{p}
    +
    \boldsymbol{\varepsilon}^{i}_{e}
    -
    \boldsymbol{\varepsilon}^{s}_{e}
    \right\rangle_{\mathrm{RVE}},
    \label{eq:main_rve_inherent_strain}
\end{equation}
where superscript $i$ denotes the peak thermal-loading state, superscript $s$ denotes the final cooled state, and the angle brackets indicate spatial averaging over the representative region of the RVE. This calibration is performed offline and does not need to be repeated during each part-scale simulation unless the material system or calibration condition is changed.

During the online stage, the part-scale thermal history predicted by the B-spline solver is used to determine the maximum temperature experienced at each material location:
\begin{equation}
    T_{\max}(\mathbf{x})=\max_t T_h(\mathbf{x},t).
    \label{eq:main_peak_temperature}
\end{equation}
The peak-temperature field is converted into the local thermal severity $\chi(\mathbf{x})$, which measures the intensity of the thermal loading relative to the temperature-dependent material strength. The RVE-calibrated inherent strain is then spatially scaled according to
\begin{equation}
    \boldsymbol{\varepsilon}^{*}(\mathbf{x})
    =
    s(\mathbf{x})\boldsymbol{\varepsilon}^{*}_{\mathrm{avg}},
    \qquad
    s(\mathbf{x})
    =
    \frac{\max\!\left[\chi(\mathbf{x})-1,0\right]}{\chi_0-1},
    \label{eq:main_inherent_strain_mapping}
\end{equation}
where $s(\mathbf{x})$ is the local inherent-strain scaling factor and $\chi_0$ is the thermal severity associated with the RVE calibration condition. In this manner, the part-scale thermal solution determines the spatial distribution of the inherent strain, while the offline RVE simulation provides its representative tensorial form.

The mapped inherent strain is finally imposed as an eigenstrain in a part-scale elastic mechanical model:
\begin{equation}
    \boldsymbol{\sigma}
    =
    \mathbb{C}:
    \left[
    \boldsymbol{\varepsilon}(\mathbf{u})
    -
    \boldsymbol{\varepsilon}^{*}(\mathbf{x})
    \right],
    \qquad
    \nabla\cdot\boldsymbol{\sigma}=\mathbf{0},
    \label{eq:main_residual_stress}
\end{equation}
where $\mathbb{C}$ is the elastic stiffness tensor, $\mathbf{u}$ is the displacement field, and $\boldsymbol{\sigma}$ is the final residual stress field. Therefore, the thermal solver provides the spatial thermal history, the offline RVE simulation provides the representative inherent strain, and the online mechanical solver maps this information to the complete part-scale residual stress distribution. This approach avoids performing a full thermo-elasto-plastic mechanical analysis throughout the entire deposition history.

The equations presented here summarize the principal thermal and residual stress simulation procedures. The complete signed-distance-function formulation, thermal weak form, B-spline Galerkin discretization, time-integration scheme, RVE calibration procedure, thermal-severity definition, inherent strain mapping, and eigenstrain finite element formulation are provided in~\ref{app:mathematical_formulation}.

\section{Thermal Model Verification and Experimental Validation}
\label{sec:validation}

The thermal predictions of the proposed B-spline Galerkin model are validated against two independent AFSD studies. The first comparison considers the layer-wise peak temperatures reported by Xiao et al.~\cite{xiao2023experimental} for five-, six-, and seven-layer AA2024 thin-wall builds. The second comparison considers the experimentally measured transient temperature history reported by Hu et al.~\cite{hu2024thermal}. Together, these comparisons assess both the accumulation of heat over successive deposition layers and the local heating and cooling response at a fixed monitoring location.

\begin{figure}[h!]
    \centering
    \includegraphics[width=0.9\linewidth]{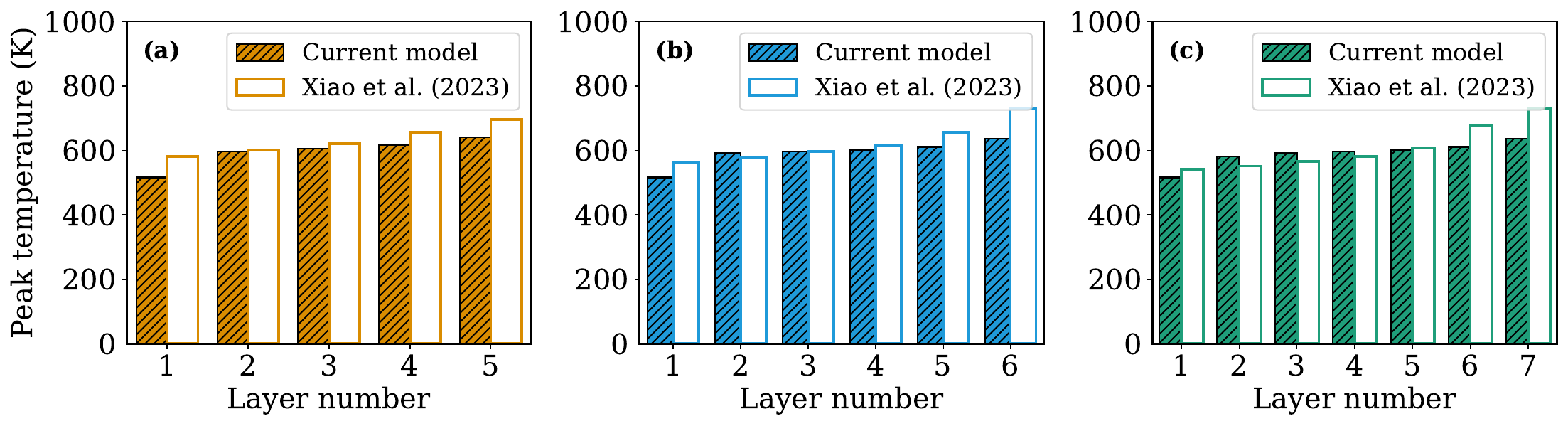}
    \caption{Layer-wise peak-temperature comparison for AA2024 thin-wall AFSD builds containing (a) five, (b) six, and (c) seven deposited layers. Predictions from the present B-spline thermal model are compared with the numerical results reported by Xiao et al.~\cite{xiao2023experimental}.}
    \label{fig:xiao_validation}
\end{figure}

Figure~\ref{fig:xiao_validation} compares the layer-wise peak temperatures predicted by the present model with the numerical results of Xiao et al.~\cite{xiao2023experimental}. The proposed model reproduces the principal trend observed in all three builds. The temperature increases after the first deposition pass because the previously deposited material and substrate remain above ambient temperature when the subsequent layer begins. The agreement is strongest during the early and intermediate layers, while a larger difference is observed near the final layers of the six- and seven-layer cases. This difference is likely associated with the treatment of heat loss, inter-layer thermal contact, and heat-source distribution, which are not identical between the two models. Nevertheless, the predicted temperature range and the progressive thermal accumulation remain consistent with the reference simulation.
% \newpage
% \begin{table}[h!]
% \centering
% \caption{Approximate layer-wise peak temperatures digitized from Figure~\ref{fig:layer_peak_validation} for the present B-spline thermal model and the numerical results of Xiao et al.~\cite{xiao2023experimental}.}
% \label{tab:layer_peak_temperature_validation}
% \begin{tabular}{cccc}
% \hline
% Build configuration & Layer & Current model (K) & Xiao et al. (K) \\
% \hline
% \multirow{5}{*}{5-layer build}
% & 1 & 515 & 580 \\
% & 2 & 600 & 600 \\
% & 3 & 605 & 620 \\
% & 4 & 610 & 655 \\
% & 5 & 640 & 695 \\
% \hline
% \multirow{6}{*}{6-layer build}
% & 1 & 515 & 560 \\
% & 2 & 590 & 580 \\
% & 3 & 595 & 600 \\
% & 4 & 600 & 615 \\
% & 5 & 610 & 660 \\
% & 6 & 640 & 730 \\
% \hline
% \multirow{7}{*}{7-layer build}
% & 1 & 515 & 540 \\
% & 2 & 580 & 550 \\
% & 3 & 590 & 565 \\
% & 4 & 595 & 580 \\
% & 5 & 605 & 610 \\
% & 6 & 610 & 675 \\
% & 7 & 640 & 730 \\
% \hline
% \end{tabular}
% \end{table}

\begin{figure}[h!]
    \centering
    \includegraphics[width=0.9\linewidth]{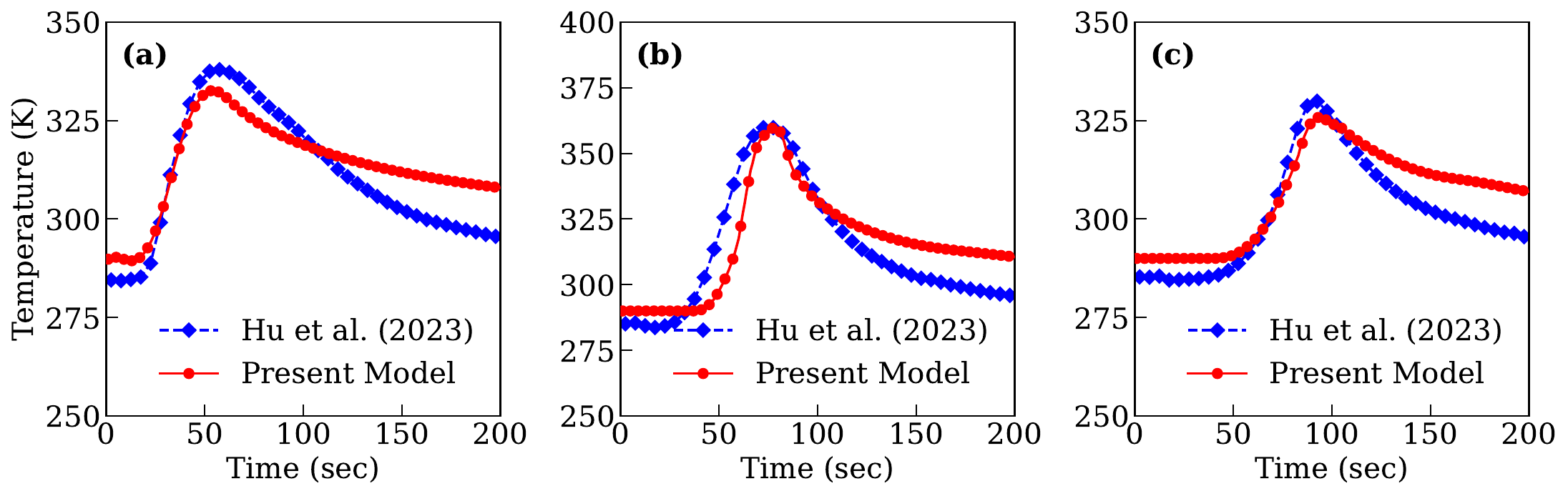}
    \caption{Transient temperature histories at three thermocouple locations along the steady-deposition path: (a) thermocouple position 4, (b) thermocouple position 5, and (c) thermocouple position 6. Predictions from the present B-spline thermal model are compared with the experimental measurements reported by Hu et al.~\cite{hu2024thermal}.}
    \label{fig:hu_validation}
\end{figure}

The transient validation shown in Figure~\ref{fig:hu_validation} provides a more direct assessment of the heating and cooling rates. The predicted response closely follows the experimental curve during the initial low-temperature period and the rapid heating stage. The model slightly underpredicts the measured peak temperature and produces a somewhat slower initial cooling response after the heat source passes the monitoring location. At later times, the predicted and measured curves converge toward the ambient temperature. The agreement indicates that the reduced-order representation retains the dominant transient conduction and surface-loss mechanisms despite using substantially fewer spatial unknowns than a conventional body-fitted finite element model.

The remaining differences for Figure~\ref{fig:hu_validation} should not be interpreted solely as reduced-order discretization error. They may also arise from uncertainty in frictional heat generation, contact conductance, local material properties, temperature-dependent boundary conditions, and the precise experimental thermocouple position. Based on these comparisons, the thermal ROM is considered sufficiently accurate for the part-scale temperature and inherent strain calculations performed in the following sections.

\section{Problem Definition and Formulation}
\label{sec:problem_formulation}

Two AFSD build configurations are considered in this study. The four-layer thin-wall build is used for validation against published experimental and finite element results, while the 20-layer hollow cylindrical build is used to evaluate the computational scalability and geometry-handling capability of the proposed framework. For each configuration, the part-scale thermal simulation is performed first, after which the predicted temperature history is transferred to the inherent strain-based mechanical model for residual stress prediction.

\subsection{Four-Layer Thin-Wall Benchmark}
\label{subsec:thinwall_setup}

\subsubsection{Geometry and Thermal Process Conditions}
\label{subsubsec:thinwall_thermal_setup}

The thermal model considers a four-layer AFSD thin-wall build matched to the experimental and finite element benchmark reported by Ruvo et al.~\cite{ruvo2026surface}. A rectangular substrate measuring $200\times150$~mm in plan and $5$~mm in thickness is used. A straight wall with a width of $30$~mm is deposited along the substrate in four successive passes. Each deposited layer is $3$~mm high, resulting in a total deposited height of $12$~mm and an overall substrate-plus-wall height of $17$~mm.

A fixed Cartesian coordinate system is defined with $x$ aligned with the tool-traverse direction, $y$ across the substrate width, and $z$ through the build height. The computational domain is therefore defined by $x\in[0,200]$~mm, $y\in[-75,75]$~mm, and $z\in[0,17]$~mm, with the deposited wall centered at $y=0$. The active material domain $\Omega_m(t)$ contains the substrate and the portion of the wall deposited up to time $t$. As each layer is added, the smooth material indicator $\eta(\mathbf{x},t)$ is updated while the Cartesian background grid remains unchanged.

\begin{figure}[h!]
\centering
\includegraphics[width=0.95\linewidth]{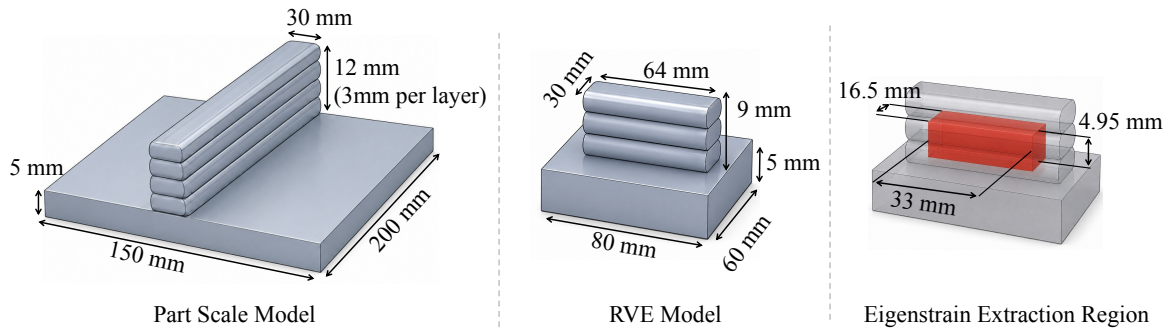}
\caption{Geometries and dimensions of the four-layer AA2024 thin-wall build and the representative volume element used for inherent strain calibration.}
\label{fig:geometry}
\end{figure}

The tool follows a straight path along the longitudinal direction of the deposited wall. A traverse speed of $v=100$~mm/min and an inter-layer dwell time of $t_{\mathrm{dwell}}=60$~s are used for the baseline simulation. The duration of one complete layer cycle is

\begin{equation}
    t_{\mathrm{cycle}}
    =
    \frac{L_x}{v}
    +
    t_{\mathrm{dwell}},
    \label{eq:thinwall_cycle_time}
\end{equation}

and the nominal end time of layer $\ell$ is given by $t_\ell=\ell t_{\mathrm{cycle}}$ for $\ell=1,\ldots,4$. The thermal-process parameters are summarized in Table~\ref{tab:thinwall_thermal_params}.

\begin{table}[h!]
\centering
\caption{Geometry and process parameters used for the thermal simulation of the four-layer thin-wall AFSD build.}
\label{tab:thinwall_thermal_params}
\begin{tabular}{lcc}
\toprule
Parameter & Symbol & Value \\
\midrule
Tool rotational speed & $\omega$ & 400, 600, and 800 rpm \\
Applied pressure & $P$ & 5, 10, and 15 MPa \\
Tool traverse speed & $v$ & 100 mm/min \\
Inter-layer dwell time & $t_{\mathrm{dwell}}$ & 60--140 s \\
Friction coefficient & $\mu_0$ & 0.3 \\
Tool pin radius & $r_p$ & 5 mm \\
Tool shoulder radius & $r_s$ & 15 mm \\
\midrule
Deposition-path length & $L_x$ & 200 mm \\
Wall width & $w_t$ & 30 mm \\
Layer height & $h_{\mathrm{layer}}$ & 3 mm \\
Number of deposited layers & $N_L$ & 4 \\
Substrate dimensions & $L_x\times L_y$ & $200\times150$ mm \\
Substrate thickness & $H_{\mathrm{sub}}$ & 5 mm \\
\bottomrule
\end{tabular}
\end{table}

The part is initialized at a uniform ambient temperature,

\begin{equation}
    T(\mathbf{x},0)
    =
    T_{\mathrm{amb}}
    =
    298~\mathrm{K},
    \qquad
    \mathbf{x}\in\Omega_m(0).
    \label{eq:thinwall_initial_temperature}
\end{equation}

A convection coefficient of $h_{\mathrm{air}}=20$~W/(m$^2$K) is applied on surfaces exposed to ambient air. The lower substrate surface is assigned $h_{\mathrm{back}}=1000$~W/(m$^2$K) to represent enhanced heat transfer into the backing fixture. Radiative heat loss is applied on the exposed surfaces using an emissivity of $\epsilon_{\mathrm{rad}}=0.7$ and a far-field temperature of $T_\infty=298$~K. The representative transient temperature fields are presented for $\omega=600$~rpm, $P=10$~MPa, and $t_{\mathrm{dwell}}=60$~s, while the complete parameter ranges are used to study the effects of rotational speed, applied pressure, and dwell time on the thermal cycle.

\subsubsection{RVE Calibration and Part-Scale Mechanical Model}
\label{subsubsec:thinwall_mechanical_setup}

The RVE shown in Figure~\ref{fig:geometry} provides the representative inherent strain used in the part-scale mechanical model. It consists of a three-layer AA2024 deposit measuring $64$ mm$\times30$~mm in plan on an $80$ mm $\times60$ mm $\times5$~mm substrate. Each deposited layer is $3$~mm high, giving a deposited height of $9$~mm and a total substrate-plus-deposit height of $14$~mm. The RVE is discretized using structured eight-node hexahedral (HEX8) elements, with an element size of $1.5$~mm in the deposited region and $3.0$~mm in the substrate, resulting in approximately $5{,}200$ elements. The lower substrate surface is fixed to represent the backing fixture, while all remaining external surfaces are traction-free.

The RVE calibration is performed using sequential thermal and thermo-elasto-plastic finite element analyses. The transient temperature history obtained from the thermal analysis drives the mechanical model with temperature-dependent elasticity, thermal expansion, and J2 plasticity. The local inherent strain is evaluated using the formulation introduced in Eq.~\eqref{eq:main_rve_inherent_strain}. To reduce the influence of free surfaces and the substrate--deposit interface, the representative inherent strain is extracted from a central region of the three-layer deposit following~\cite{dong2021new}. The extraction region measures $33$ mm $\times16.5$ mm $\times4.95$~mm in the longitudinal, transverse, and build directions, respectively, and is centered within the deposited material.

The representative inherent strain tensor is obtained by volume-averaging the local inherent strain over this extraction region:
\begin{equation}
    \boldsymbol{\varepsilon}^{*}_{\mathrm{avg}}
    =
    \frac{
        \displaystyle
        \sum_{k\in\mathcal{R}}
        \sum_{m}
        J_{km}\,w_m\,\boldsymbol{\varepsilon}^{*}_{km}
    }{
        \displaystyle
        \sum_{k\in\mathcal{R}}
        \sum_{m}
        J_{km}\,w_m
    }.
    \label{eq:rve_average}
\end{equation}
Here, $\boldsymbol{\varepsilon}^{*}_{\mathrm{avg}}$ is the volume-averaged inherent strain tensor used for the part-scale model, $\mathcal{R}$ denotes the set of finite elements contained within the extraction region, $k$ denotes an element in $\mathcal{R}$, and $m$ denotes a numerical integration point within that element. The tensor $\boldsymbol{\varepsilon}^{*}_{km}$ is the local inherent strain evaluated at integration point $m$ of element $k$, $w_m$ is the corresponding quadrature weight, and $J_{km}$ is the determinant of the element mapping Jacobian at that integration point. Thus, $J_{km}w_m$ represents the physical volume associated with the integration point, and Eq.~\eqref{eq:rve_average} computes a physical-volume-weighted average of the inherent strain over the selected RVE region. The $11$, $22$, and $33$ components of $\boldsymbol{\varepsilon}^{*}_{\mathrm{avg}}$ correspond to the longitudinal, transverse, and build directions, respectively.

The same offline RVE simulation also provides the reference thermal severity $\chi_0$ associated with the calibration condition. The quantities $\boldsymbol{\varepsilon}^{*}_{\mathrm{avg}}$ and $\chi_0$ are stored after the RVE calibration and subsequently used in the part-scale inherent-strain mapping defined in Eq.~\eqref{eq:main_inherent_strain_mapping}. The resulting spatially varying eigenstrain field is then imposed in the final elastic equilibrium problem described in Section~\ref{sec:computational_methodology}.

\subsection{Twenty-Layer Hollow-Cylinder Demonstration}
\label{subsec:cylinder_setup}

\subsubsection{Geometry and Thermal Process Conditions}
\label{subsubsec:cylinder_thermal_setup}

The second thermal model considers a 20-layer hollow cylindrical wall deposited on a rectangular substrate. The substrate measures $180\times180$~mm in plan and has a thickness of $5$~mm. The cylindrical wall has an outer diameter of $120$~mm, an inner diameter of $60$~mm, a wall thickness of $30$~mm, and a total deposited height of $60$~mm. The wall is constructed from 20 layers, each with a height of $3$~mm. The tool follows the circular wall centerline with a radius of $R=45$~mm. Unlike the straight thin-wall case, the deposited geometry evolves around a closed circumferential path. Progressive material activation is represented using the same signed-distance-function and fixed-grid formulation used for the thin-wall build.

\begin{figure}[h!]
\centering
\includegraphics[width=1\textwidth]{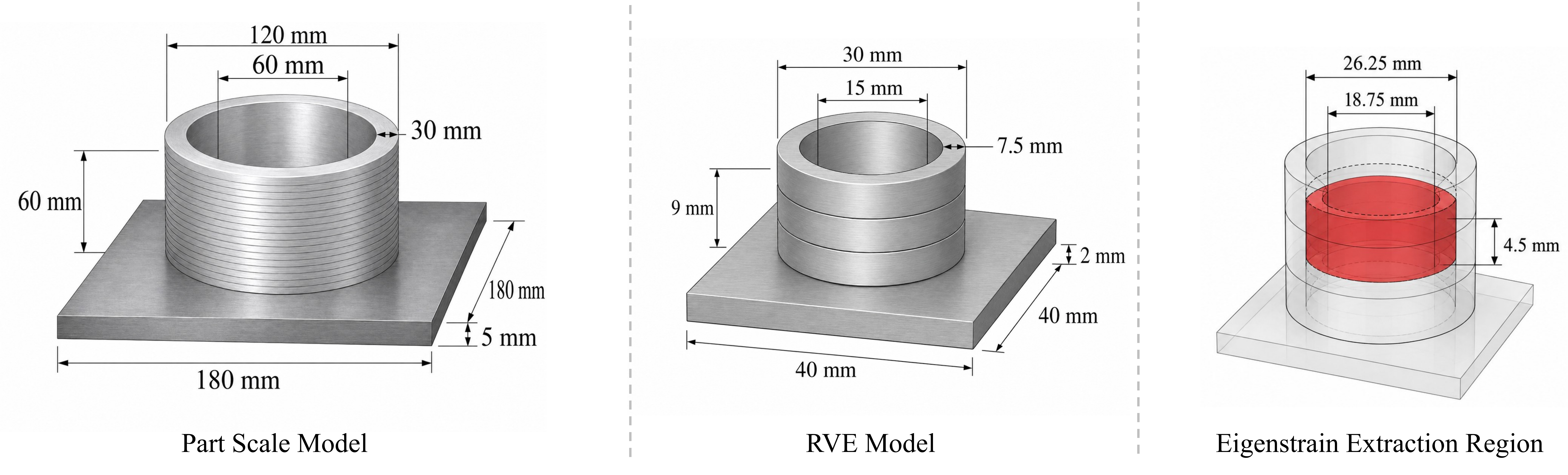}
\caption{Geometries and dimensions of the 20-layer hollow cylindrical AA2024 build, the cylindrical RVE used for inherent strain calibration, and the annular region used for eigenstrain extraction.}
\label{fig:cylinder_geometry}
\end{figure}

The thermal process parameters are summarized in Table~\ref{tab:cylinder_thermal_params}. One complete circumferential sweep requires approximately $169.6$~s at the prescribed traverse speed. Including the $60$~s dwell period, the complete layer-cycle time is $229.6$~s and the total simulated build duration is approximately $4,593$~s.

\begin{table}[h!]
\centering
\caption{Geometry and process parameters used for the thermal simulation of the 20-layer hollow cylindrical AFSD build.}
\label{tab:cylinder_thermal_params}
\begin{tabular}{lcc}
\toprule
Parameter & Symbol & Value \\
\midrule
Tool rotational speed & $\omega$ & 400, 600 and 800 rpm \\
Applied pressure & $P$ & 10 MPa \\
Friction coefficient & $\mu_0$ & 0.3 \\
Tool traverse speed & $v$ & 100 mm/min \\
Inter-layer dwell time & $t_{\mathrm{dwell}}$ & 60 s \\
Tool pin radius & $r_p$ & 5 mm \\
Tool shoulder radius & $r_s$ & 15 mm \\
\midrule
Wall centerline radius & $R$ & 45 mm \\
Wall thickness & $w_t$ & 30 mm \\
Layer height & $h_{\mathrm{layer}}$ & 3 mm \\
Number of deposited layers & $N_L$ & 20 \\
Substrate dimensions & $L_x\times L_y$ & $180\times180$ mm \\
Substrate thickness & $H_{\mathrm{sub}}$ & 5 mm \\
\midrule
Circumferential sweep time & $t_{\mathrm{sweep}}$ & 169.6 s \\
Layer-cycle time & $t_{\mathrm{cycle}}$ & 229.6 s \\
Total build time & $t_{\mathrm{total}}$ & 4,593 s \\
\bottomrule
\end{tabular}
\end{table}

The cylindrical model is initialized at $T_{\mathrm{amb}}=298$~K and uses the same convective, radiative, and backing-fixture thermal boundary conditions as the thin-wall build. The same temperature-dependent AA2024 thermal properties are also used.

\subsubsection{Cylindrical RVE Calibration and Mechanical Model}
\label{subsubsec:cylinder_mechanical_setup}

The cylindrical RVE shown in Figure~\ref{fig:cylinder_geometry} is used to calibrate the inherent strain for the hollow cylindrical build following the same offline thermo-elasto-plastic procedure described in Section~\ref{subsubsec:thinwall_mechanical_setup}. The RVE consists of a three-layer hollow cylindrical AA2024 deposit on a $40\times40$~mm substrate with a thickness of $2$~mm. The deposited cylinder has an outer diameter of $30$~mm, an inner diameter of $15$~mm, and a wall thickness of $7.5$~mm. Each deposited layer is $3$~mm high, resulting in a total deposited height of $9$~mm.

The RVE is discretized using structured Hex8 voxel elements. The lower surface of the substrate is fixed to represent the backing fixture, while all remaining surfaces are traction-free. The temperature history obtained from the finite element thermal simulation is applied to the thermo-elasto-plastic mechanical model using the same temperature-dependent material behavior adopted for the thin-wall RVE. To reduce the influence of the free surfaces and the substrate--deposit interface, the inherent strain is volume-averaged over a central annular region within the cylindrical deposit. The extraction region occupies the central $50\%$ of the wall thickness and the central $50\%$ of the deposited height while extending around the full circumference. It therefore has an outer diameter of $26.25$~mm, an inner diameter of $18.75$~mm, and a height of $4.5$~mm. The representative inherent strain is calculated using Eq.~\eqref{eq:rve_average}. During the part-scale cylindrical analysis, the calibrated inherent strain tensor and reference thermal severity are spatially scaled using Eq.~\eqref{eq:main_inherent_strain_mapping} and imposed in the final elastic equilibrium solve.

\section{Results and Analysis}

The results are presented separately for the four-layer thin-wall build and the 20-layer hollow cylindrical build. For each configuration, the part-scale thermal response is discussed first, followed by the effects of the investigated process parameters, the RVE calibration results, the final part-scale residual stress prediction, and the computational performance of the proposed framework. All finite-element and reduced-order simulations presented in this study are performed using NVIDIA A100 GPU resources provided by Virginia Tech Advanced Research Computing (ARC) \cite{vt_arc_tinkercliffs}, ensuring a consistent computational platform for the reported runtime and speed-up comparisons.

\subsection{Analysis of a Four Layer Thin-Wall AFSD Build}
\label{sec:thin_wall}

\subsubsection{Comparison of Numerical Speedup}
\label{subsec:timing_overview}

\begin{figure}[h!]
  \centering
  \includegraphics[width=0.95\textwidth]{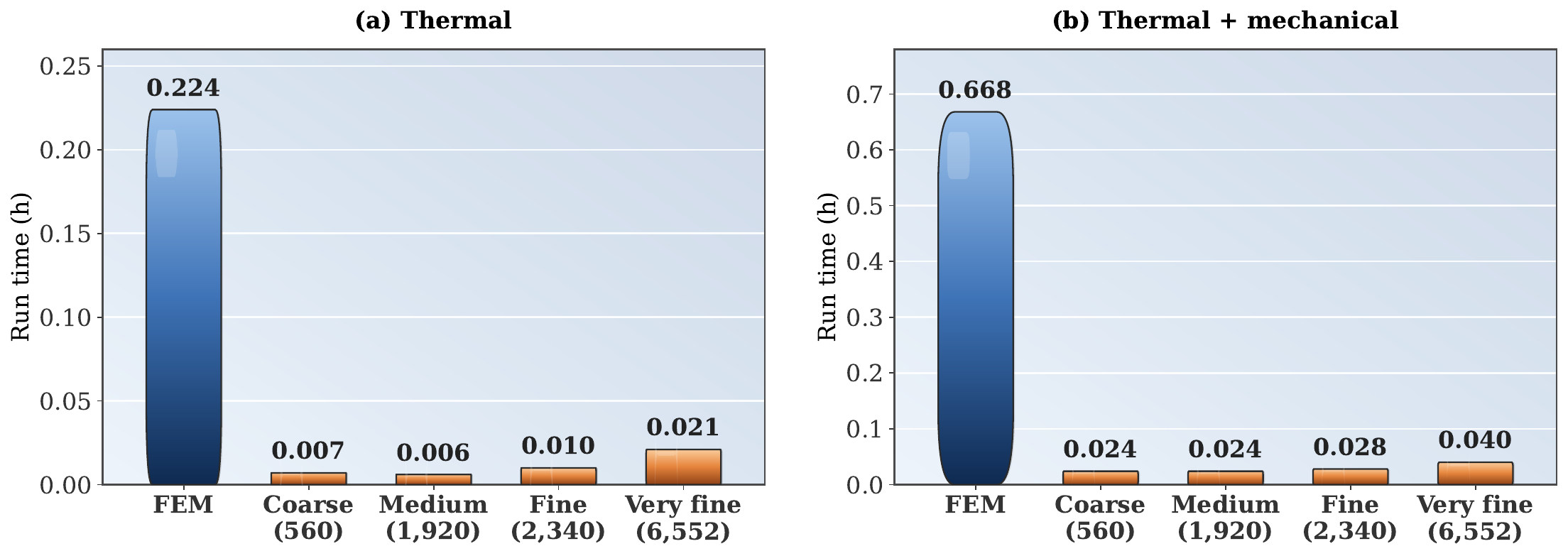}
  \caption{Solution time for the four-layer AA2024 thin-wall build: (a) thermal solve only, compared across four B-spline resolutions, and (b) combined thermal and mechanical solve. The numbers in parentheses indicate the corresponding reduced-order model DOFs. }
  \label{fig:thinwall_timing}
\end{figure}

The computational cost of the proposed framework is analysed in this section for the four-layer thin-wall build. First, the thermal solve is compared, followed by the combined thermal-plus-mechanical solution time, so that the contribution of the eigenstrain mechanical solve to the overall speed-up can be isolated. The thermal solver is benchmarked against the finite element reference for the four-layer AA2024 thin-wall build across four B-spline resolutions (Figure~\ref{fig:thinwall_timing}a). The finite element thermal simulation requires 806~s, whereas the reduced-order thermal simulations require between 21.6~s and 75.6~s, with the very fine resolution model (6,552 DOF) giving the highest runtime at 75.6~s -- an approximately 10.6-fold reduction relative to the finite element model. The medium resolution model (1,920 DOF) shows a 37.3 fold speed up. When the mechanical eigenstrain solve is included, the total solution time is reduced by a comparable margin (Figure~\ref{fig:thinwall_timing}b). The finite element thermomechanical simulation requires approximately 2406~s in total, whereas the ROM/ISM pipeline requires between 86~s and 144~s across the same four resolutions, with the very fine-resolution model completing the full thermomechanical solve in 144~s which is an approximately 17-fold reduction relative to the finite element model.

%\subsubsection{Comparison of Thermal Distribution}
\subsubsection{Transient Thermal Evolution}
\label{subsec:temp_distribution}

\begin{figure}[h!]
  \centering
  \includegraphics[width=0.95\textwidth]{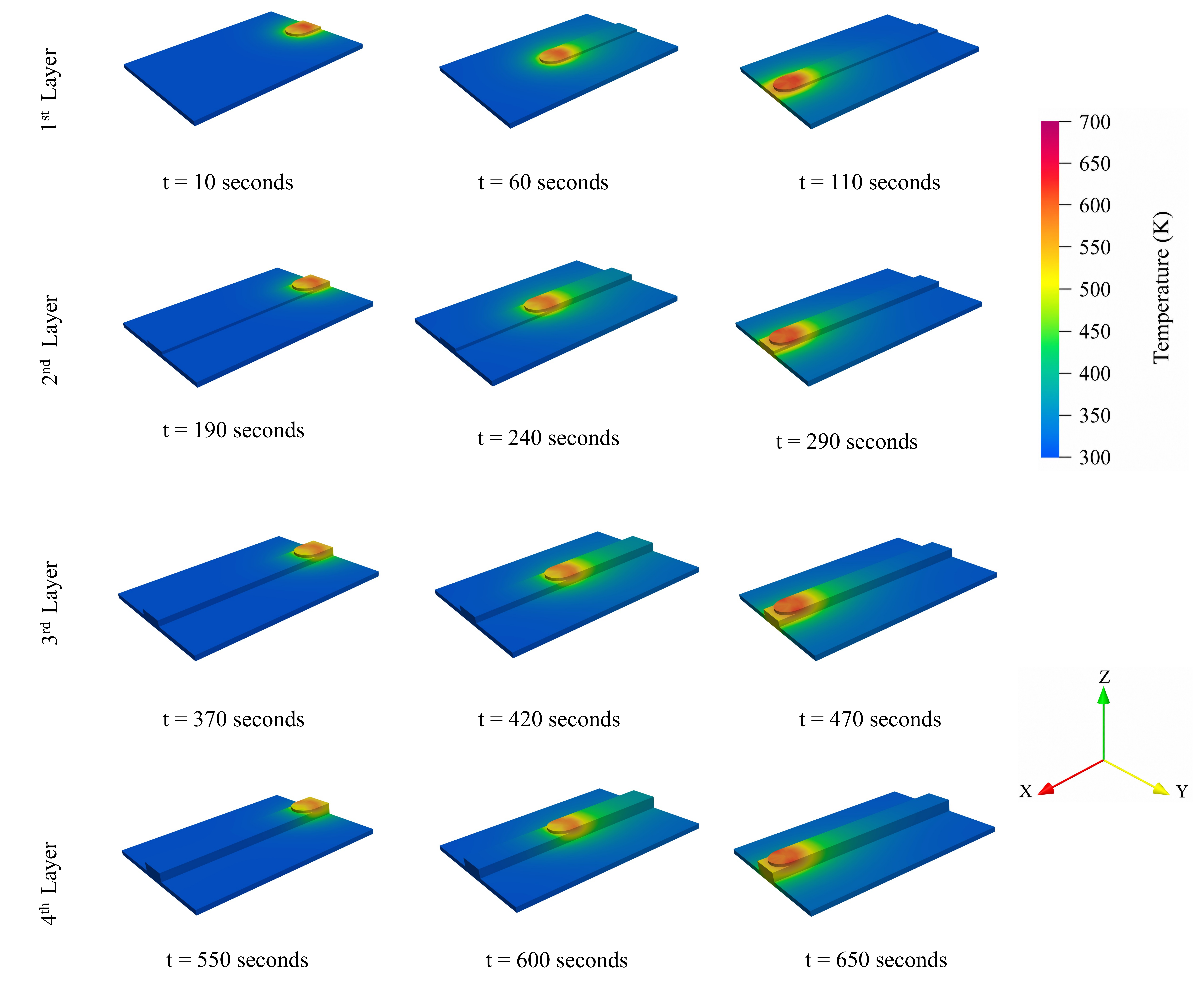}
  \caption{Transient temperature distribution during the four-layer AFSD thin-wall build predicted by the B-spline thermal ROM. Three snapshots are shown for each layer, corresponding approximately to the beginning, middle, and end of the deposition pass.}
  \label{fig:transient_temperature}
\end{figure}

The transient temperature field predicted by the B-spline thermal ROM is shown in Figure~\ref{fig:transient_temperature} for the four-layer AA2024 thin-wall build. The simulation is performed at a tool rotational speed of 600~rpm, a traverse speed of 100~mm/min, an applied pressure of 10~MPa, and an inter-layer dwell time of 60~s. Three snapshots are shown for each layer, corresponding to the early, intermediate, and final portions of the deposition pass. During each pass, the highest temperature remains localized beneath and immediately behind the moving tool. At the beginning of a layer, the heat source interacts with relatively cool material and a large temperature gradient forms between the active deposition zone and the surrounding substrate. As the tool progresses along the wall, heat accumulates in the previously traversed region and the thermal field becomes elongated along the deposition direction. The wake behind the tool therefore remains elevated in temperature even after the local heat input has moved forward.

The previously deposited layers also influence the temperature distribution of the subsequent passes. Their elevated initial temperature reduces the thermal gradient near the active layer, while the substrate continues to act as the principal heat sink. Consequently, the thermal field progressively extends through the deposited wall and into the substrate as the build height increases. The 60~s dwell period permits partial cooling between passes but does not return the complete part to ambient temperature, producing a repeated heating and cooling response over the four-layer build.

\subsubsection{Through-Thickness and Longitudinal Temperature Fields} \label{subsec:sectional_temperature} 
\begin{figure}[h!]
\centering
\includegraphics[width=0.95\textwidth]{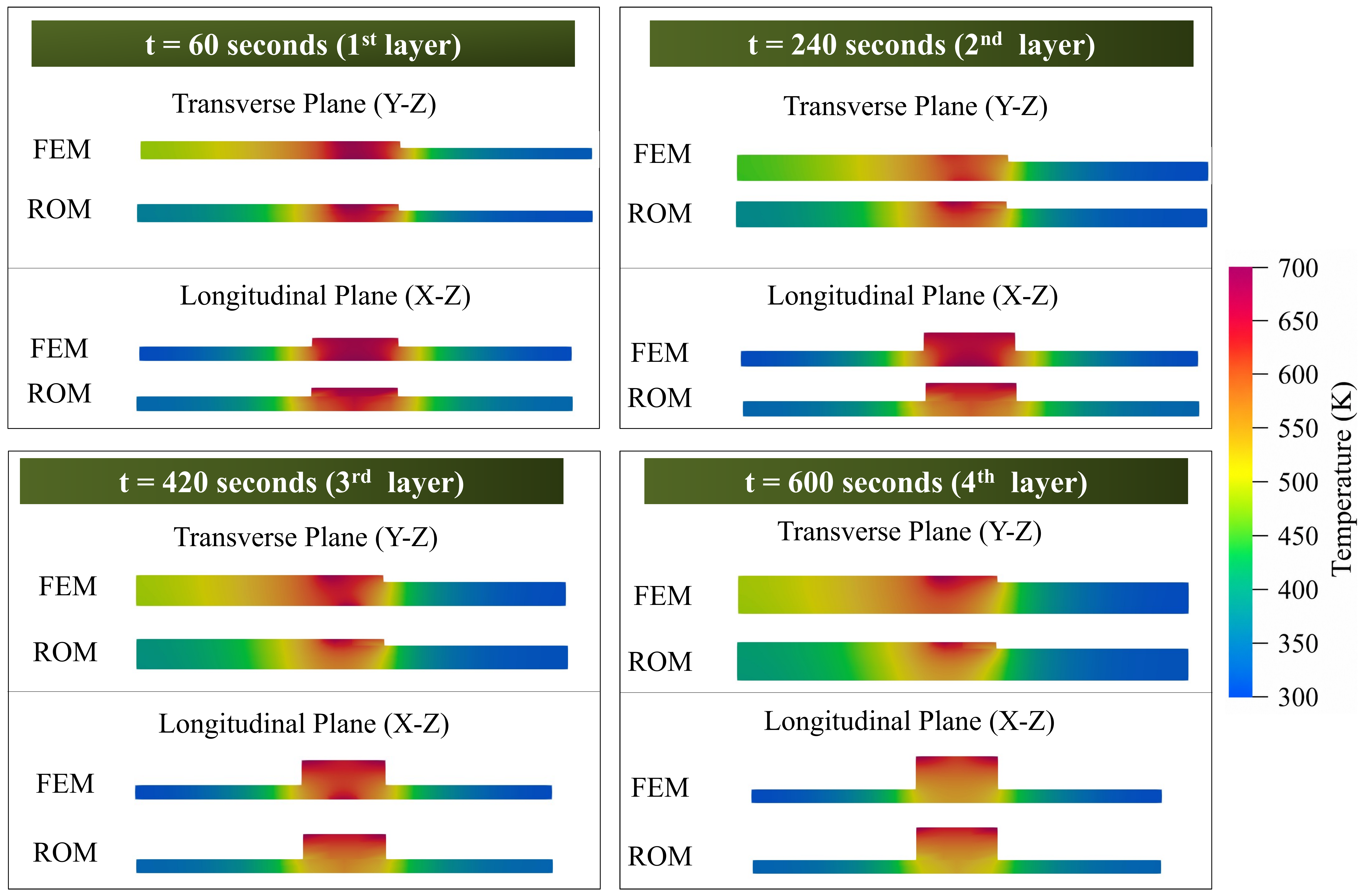}
\caption{Comparison of FEM and B-spline ROM temperature distributions on transverse ($Y$--$Z$) and longitudinal ($X$--$Z$) section planes at the midpoint of the first, second, third, and fourth deposition passes.}
\label{fig:sectional_temperature}
\end{figure}

Figure~\ref{fig:sectional_temperature} compares the temperature fields predicted by the B-spline ROM and the finite element model on transverse ($Y$--$Z$) and longitudinal ($X$--$Z$) section planes at the midpoint of each deposition pass. The ROM closely reproduces the spatial temperature distributions obtained from the FEM throughout all four layers, including the location of the peak-temperature region and the extent of heat diffusion into the surrounding material. The transverse $Y$--$Z$ sections show that the highest temperatures remain concentrated near the upper deposited surface in the vicinity of the active deposition region. Heat is conducted downward through the deposited wall and into the substrate, producing a progressively broader lower-temperature region beneath the active layer. As the build height increases from the first to the fourth layer, the high-temperature region moves upward with the active deposition surface while previously deposited material continues to conduct heat toward the substrate.

The longitudinal $X$--$Z$ sections further demonstrate the asymmetric thermal field generated by the moving heat source. A comparatively steep temperature gradient is observed ahead of the active deposition region, whereas an extended elevated-temperature wake develops behind it along the previously traversed portion of the wall. This leading--trailing asymmetry is reproduced by the ROM for each deposition layer. Minor differences between the two solutions are primarily visible in the smoothness and spatial extent of the intermediate-temperature gradients, while the position and overall dimensions of the high-temperature region remain in close agreement. The accurate recovery of these spatial temperature fields is particularly important for the subsequent inherent strain analysis, where the local eigenstrain magnitude is determined from the peak-temperature history throughout the deposited part.

%\subsection{Effect of Process Parameters on Thermal Cycle}

\subsubsection{Pressure Dependence of the Thermal Cycle}

\begin{figure}[h!]
    \centering
    \includegraphics[width=1\linewidth]{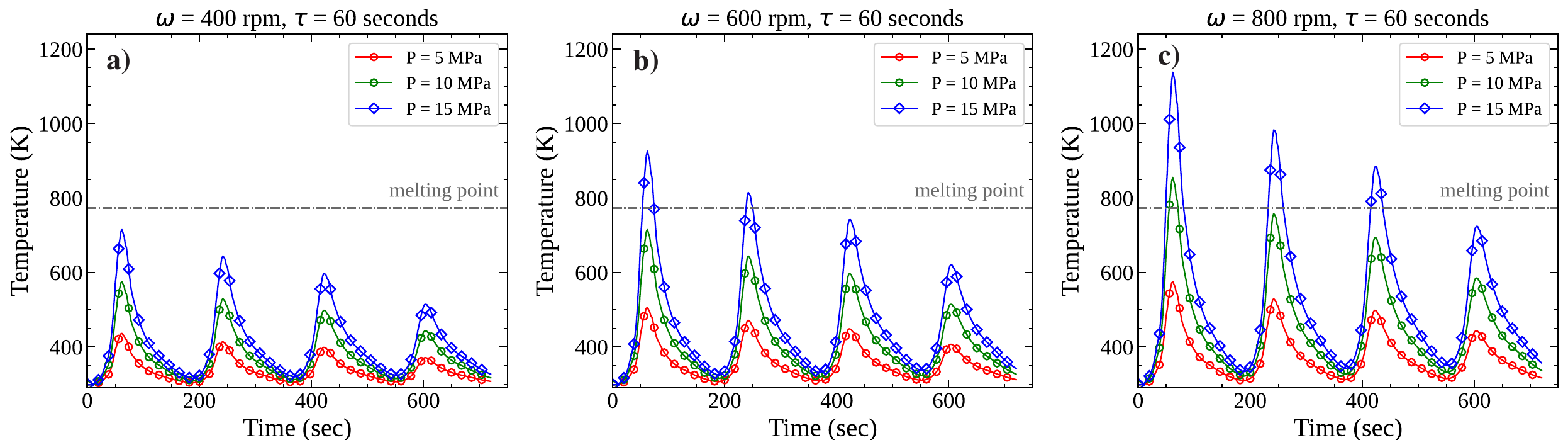}
    \caption{Predicted temperature history at a fixed monitoring point for tool rotational speeds of (a) 400, (b) 600, and (c) 800~rpm, each at applied pressures of 5, 10, and 15~MPa and a fixed dwell time of 60~s.}
    \label{fig:thermal_vs_pressure}
\end{figure}

Figure~\ref{fig:thermal_vs_pressure} shows the predicted temperature history at a fixed monitoring location on the wall centerline for tool rotational speeds of 400, 600, and 800~rpm, each evaluated at applied pressures of 5, 10, and 15~MPa with the inter-layer dwell fixed at 60~s. The monitoring point is located at mid-length along the tool-traverse direction, on the wall centerline, at the top of the substrate ($x=100$~mm, $y=0$~mm, $z=5$~mm in the coordinate convention of Section~\ref{subsubsec:thinwall_thermal_setup}), coinciding with the substrate--Layer~1 interface. At a given rotational speed, increasing the applied pressure raises the peak temperature in every pass, and the sensitivity to pressure grows with rotational speed; at 800~rpm and 15~MPa the first-layer peak exceeds the AA2024 solidus temperature, placing this combination outside the intended solid-state process window. Within each curve, the peak temperature registered at the monitoring point decreases from the first to the fourth pass, reflecting the increasing distance between the tool and the fixed monitoring location as the build height increases, rather than a reduction in the total heat generated.

\subsubsection{Dwell-Time Dependence of the Thermal Cycle}

\begin{figure}[h!]
    \centering
    \includegraphics[width=1\linewidth]{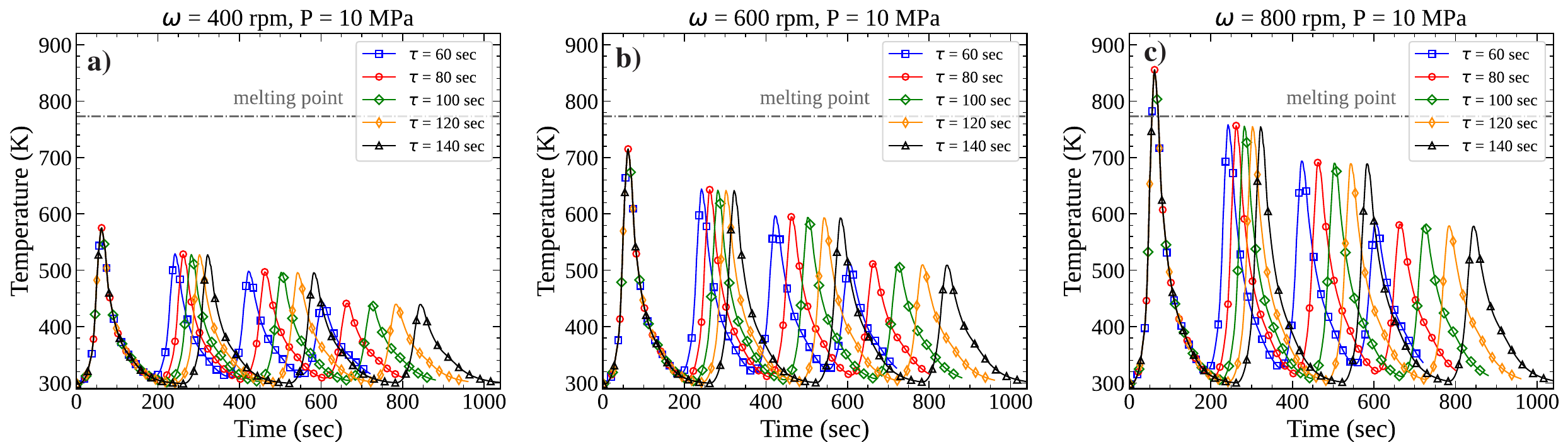}
    \caption{Predicted temperature history at a fixed monitoring point for tool rotational speeds of (a) 400, (b) 600, and (c) 800~rpm at a fixed applied pressure of 10~MPa, for inter-layer dwell times of 60, 80, 100, 120, and 140~s.}
    \label{fig:thermal_vs_dwell}
\end{figure}

Figure~\ref{fig:thermal_vs_dwell} presents the same monitoring-point thermal cycle for inter-layer dwell times of 60, 80, 100, 120, and 140~s at a fixed applied pressure of 10~MPa. Because the dwell is only applied between passes, the first thermal excursion is nearly unaffected by dwell time. From the second pass onward, longer dwell times delay the onset of each subsequent cycle, consistent with $t_{\mathrm{cycle}} = L_x/v + t_{\mathrm{dwell}}$, and reduce the peak temperature reached in each pass by allowing more inter-pass cooling. This effect is most pronounced at 800~rpm, where the fourth-layer peak drops by several tens of kelvin as the dwell increases from 60 to 140~s. Together, Figures~\ref{fig:thermal_vs_pressure} and~\ref{fig:thermal_vs_dwell} indicate that applied pressure primarily governs the magnitude of the thermal input, while dwell time governs the degree of inter-pass cooling carried forward into subsequent layers.

\subsubsection{Process-Dependent residual stress Prediction}
\label{subsec:residual_stress_results}

\begin{figure}[h!]
    \centering
    \includegraphics[width=1\linewidth]{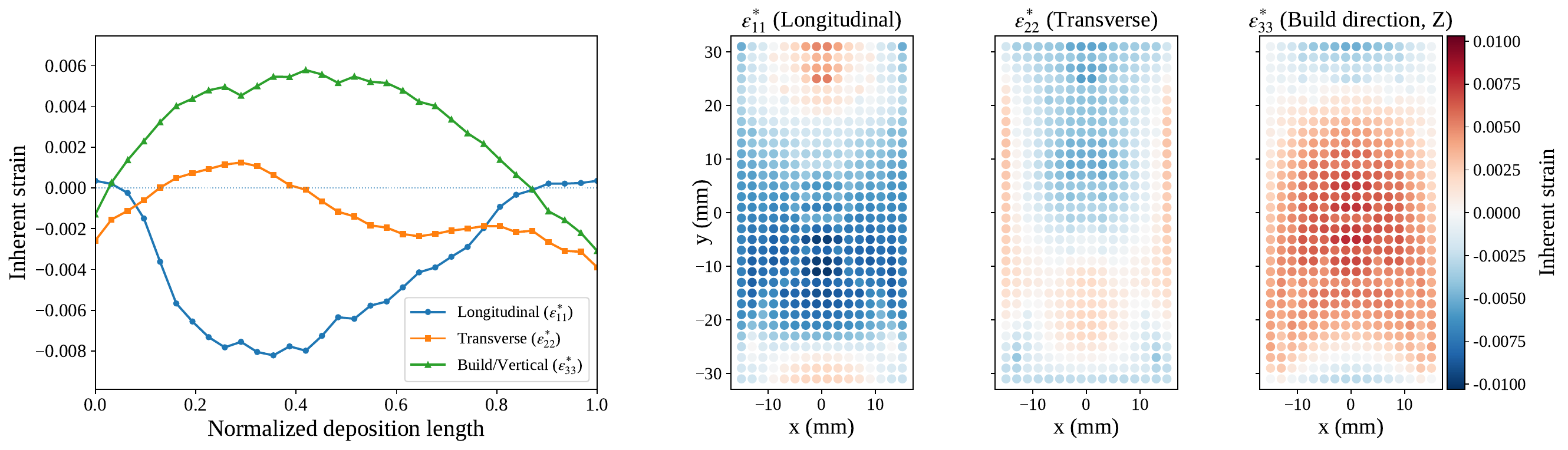}
    \caption{RVE-extracted elemental inherent strain for the three-layer thin-wall representative volume element: (left) longitudinal, transverse, and build-direction inherent strain components plotted along the normalized deposition length of the first layer of deposition, and (right) the corresponding spatial distribution of each component over the deposited region.}
    \label{fig:thinwall_inherent_strain}
\end{figure}

The RVE-extracted inherent strain field used to calibrate the thin-wall part-scale eigenstrain model is shown in Figure~\ref{fig:thinwall_inherent_strain}. The longitudinal component $\varepsilon^{*}_{11}$ is negative over most of the deposited length, reflecting the dominant contraction generated along the tool-traverse direction, while the build-direction component $\varepsilon^{*}_{33}$ remains positive throughout, consistent with the expected Poisson-type coupling between the in-plane contraction and the through-thickness expansion of the deposited layer. The transverse component $\varepsilon^{*}_{22}$ remains comparatively small in magnitude relative to $\varepsilon^{*}_{11}$ and $\varepsilon^{*}_{33}$, indicating a weaker inherent strain contribution in the direction perpendicular to the tool path. Localized departures from this trend are visible near both ends of the deposition length, coinciding with the plunge and retraction regions of the tool path, where the spatial maps show a concentrated region of elevated $\varepsilon^{*}_{11}$ magnitude near one edge of the deposit that is not present in the interior of the wall. The magnitude of the extracted inherent strain, on the order of $10^{-3}$, is approximately an order of magnitude smaller than the values reported for fusion-based directed energy deposition and laser powder bed fusion processes~\cite{liang2018modified,liang2020homogenization}, where inherent strain is generated primarily by rapid solidification shrinkage of a melt pool and the associated large thermal gradients accompanying melting and resolidification. Because AFSD is a solid-state process that thermomechanically deforms feedstock without melting~\cite{yu2018nonbeam,yu2021deformation}, no such solidification-driven strain is present; instead, the inherent strain arises from severe plastic shear deformation imposed by the rotating tool and from thermal contraction following a peak process temperature that remains below the solidus~\cite{gotawala2023material,yakubov2024evolution}, which is consistent with the comparatively lower strain magnitudes observed here relative to fusion-based inherent strain calibrations. Away from the plunge and retraction regions, the spatial distribution of all three components is comparatively uniform along the deposition direction, supporting the use of a single volume-averaged inherent strain tensor $\boldsymbol{\varepsilon}^{*}_{\mathrm{avg}}$ for the part-scale eigenstrain mapping described in ~\ref{subsec:rve_calibration}. 

The volume-averaged inherent strain tensor and reference thermal severity obtained from the thin-wall RVE calibration are summarized in Table~\ref{tab:thinwall_rve_values}. These values are computed using Eq.~\eqref{eq:rve_average} over the representative region defined in Section~\ref{subsubsec:thinwall_mechanical_setup} and are used directly, without modification, in the part-scale eigenstrain mapping of Eq.~\eqref{eq:main_inherent_strain_mapping} for all thin-wall residual stress predictions reported in this section.

\begin{table}[h!]
\centering
\caption{Volume-averaged inherent strain components extracted from the thin-wall RVE calibration.}
\label{tab:thinwall_rve_values}
\begin{tabular}{lc}
\toprule
\textbf{Quantity} & \textbf{Value} \\
\midrule
$\varepsilon^{*}_{11}$ (longitudinal) & $-5.109\times10^{-3}$ \\
$\varepsilon^{*}_{22}$ (transverse)   & $+1.504\times10^{-3}$ \\
$\varepsilon^{*}_{33}$ (build)        & $+7.320\times10^{-3}$ \\
\end{tabular}
\end{table}

\begin{figure}[h!]
    \centering
    \includegraphics[width=0.7\linewidth]{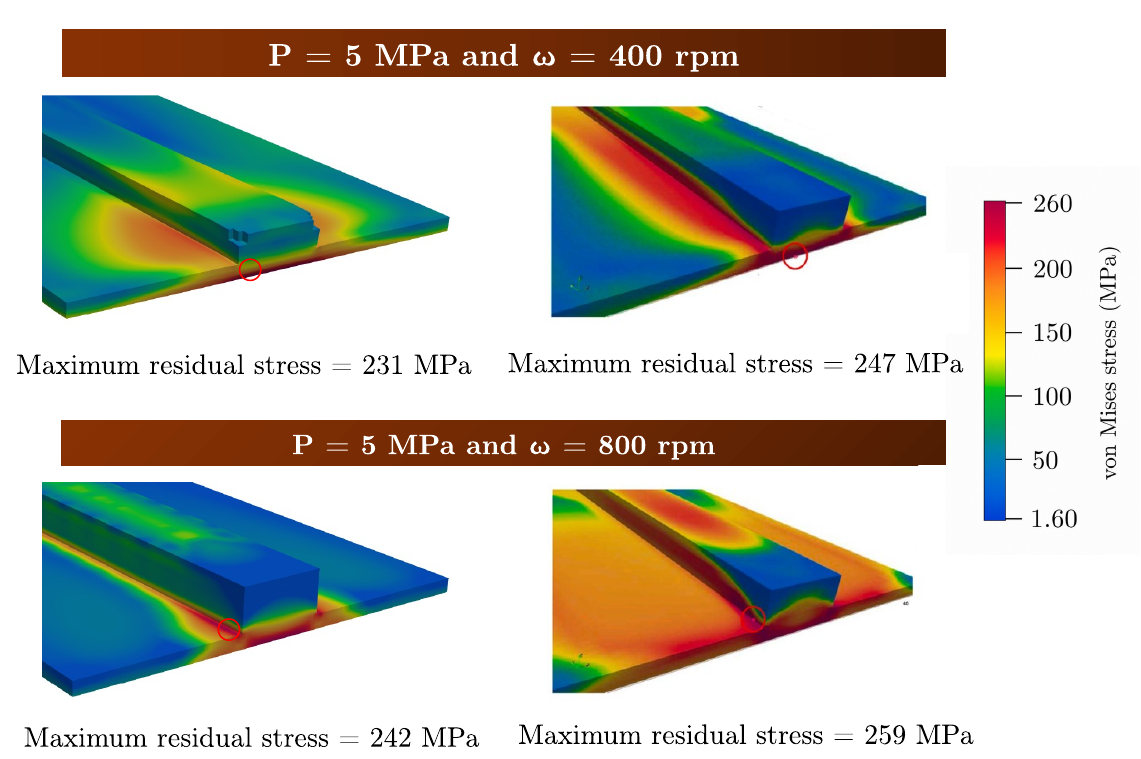}
    \caption{Comparison of the final von Mises residual stress distributions predicted by the B-spline ROM/ISM framework (left) and the finite element model (right) for tool rotational speeds of 400 rpm (top) and 800 rpm (bottom) at an applied pressure of 5 MPa.}
    \label{fig:residual_stress}
\end{figure}

The final von Mises residual stress distributions predicted by the eigenstrain solver are shown in Figure~\ref{fig:residual_stress} for tool rotational speeds of 400 and 800~rpm. For these mechanical simulations, the applied pressure is fixed at 5~MPa. The part-scale eigenstrain field is obtained by scaling the RVE-calibrated inherent strain according to the local thermal severity calculated from the maximum-temperature history. For both process conditions, the largest residual stresses are concentrated in and around the deposited wall, particularly near the wall--substrate transition. This region experiences both a strong thermal gradient during deposition and substantial mechanical constraint during cooling. The substrate away from the wall remains at a lower stress level because it undergoes a smaller thermal excursion and is less directly affected by the imposed eigenstrain.

Changing the rotational speed modifies both the magnitude and spatial extent of the residual stress field. The higher rotational speed increases the frictional heat input and enlarges the region subjected to elevated peak temperatures. As a result, a larger portion of the deposited wall receives a nonzero scaled inherent strain. The corresponding stress field becomes more spatially distributed along the deposition path and through the wall thickness. The observed response is consistent with the direct dependence of the thermal severity measure on the predicted peak temperature. Table~\ref{tab:residual_stress} summarizes the maximum residual stress predicted by the FEM model and compares it with the corresponding B-spline ROM predictions for the investigated processing conditions.

\begin{table}[ht]
\centering
\caption{Comparison of maximum residual stress predicted by the FEM model and the B-spline ROM under different rotational speeds at a constant applied pressure of $P=5$ MPa.}
\label{tab:residual_stress}
\begin{tabular}{ccc}
\hline
\textbf{Rotational Speed (rpm)} & \textbf{FEM (MPa)} & \textbf{B-spline ROM (MPa)} \\
\hline
400 & 247 &  231\\
800 & 259 &  242\\
\hline
\end{tabular}
\end{table}

The predicted field represents the final self-equilibrated stress state after deposition and cooling. It does not reproduce the complete incremental stress path that would be obtained from a full part-scale thermo-elasto-plastic analysis. Instead, the irreversible local response is calibrated once using the RVE model and transferred to the part scale as an imposed eigenstrain. This treatment preserves the primary influence of the thermal history and structural constraint while reducing the mechanical analysis to a single elastic equilibrium solve for each process condition.

\subsection{Analysis of the Twenty-Layer Hollow Cylindrical AFSD Build}
\label{sec:cylinder_results}

The 20-layer hollow cylindrical build is used to evaluate the performance of the proposed framework for a multilayer geometry involving a curved, closed deposition path. The analysis considers computational efficiency, transient thermal evolution, final residual stress development, and the sensitivity of the Layer~1 thermal history to rotational speed, applied pressure, and inter-layer dwell time.

\subsubsection{Computational Performance and Scalability}
\label{subsubsec:cylinder_timing}

\begin{figure}[h!]
\centering
\includegraphics[width=1\textwidth]{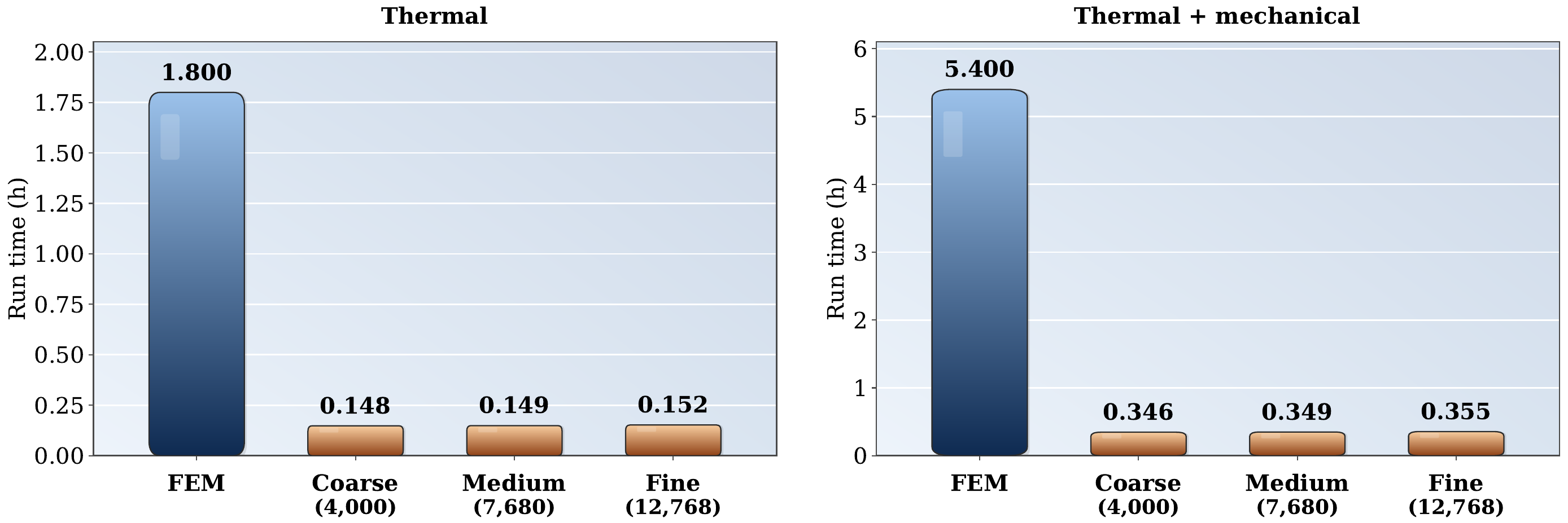}
\caption{Computational time for the 20-layer hollow cylindrical AFSD build: (a) thermal solution and (b) complete thermal and mechanical solution.}
\label{fig:cylinder_timing}
\end{figure}

Figure~\ref{fig:cylinder_timing} compares the computational cost of the full-order finite element model and the proposed reduced-order framework for the 20-layer hollow cylindrical build. The finite element thermal simulation requires 1.800~h, whereas the B-spline Galerkin ROM required only 0.148--0.152~h across the investigated resolutions, corresponding to thermal speed-ups of approximately 11.8--12.2$\times$. The computational advantage is retained when the mechanical analysis is included. The finite element mechanical solve requires an additional 3.600~h, resulting in a total thermomechanical solution time of 5.400~h. In comparison, the inherent strain mechanical solve requires only 0.198--0.203~h, corresponding to a mechanical speed-up of approximately 17.7--18.2$\times$. Including both the thermal and mechanical calculations, the complete ROM/ISM framework requires 0.346--0.355~h, compared with 5.400~h for the full finite element analysis, providing an overall speed-up of approximately 15.2--15.6$\times$. These results demonstrate that the computational advantage of the proposed framework is retained for a substantially longer deposition history and a geometrically more complex, curved deposition path.

\subsubsection{Transient Thermal Evolution}

\begin{figure}[h!]
    \centering
    \includegraphics[width=1\linewidth]{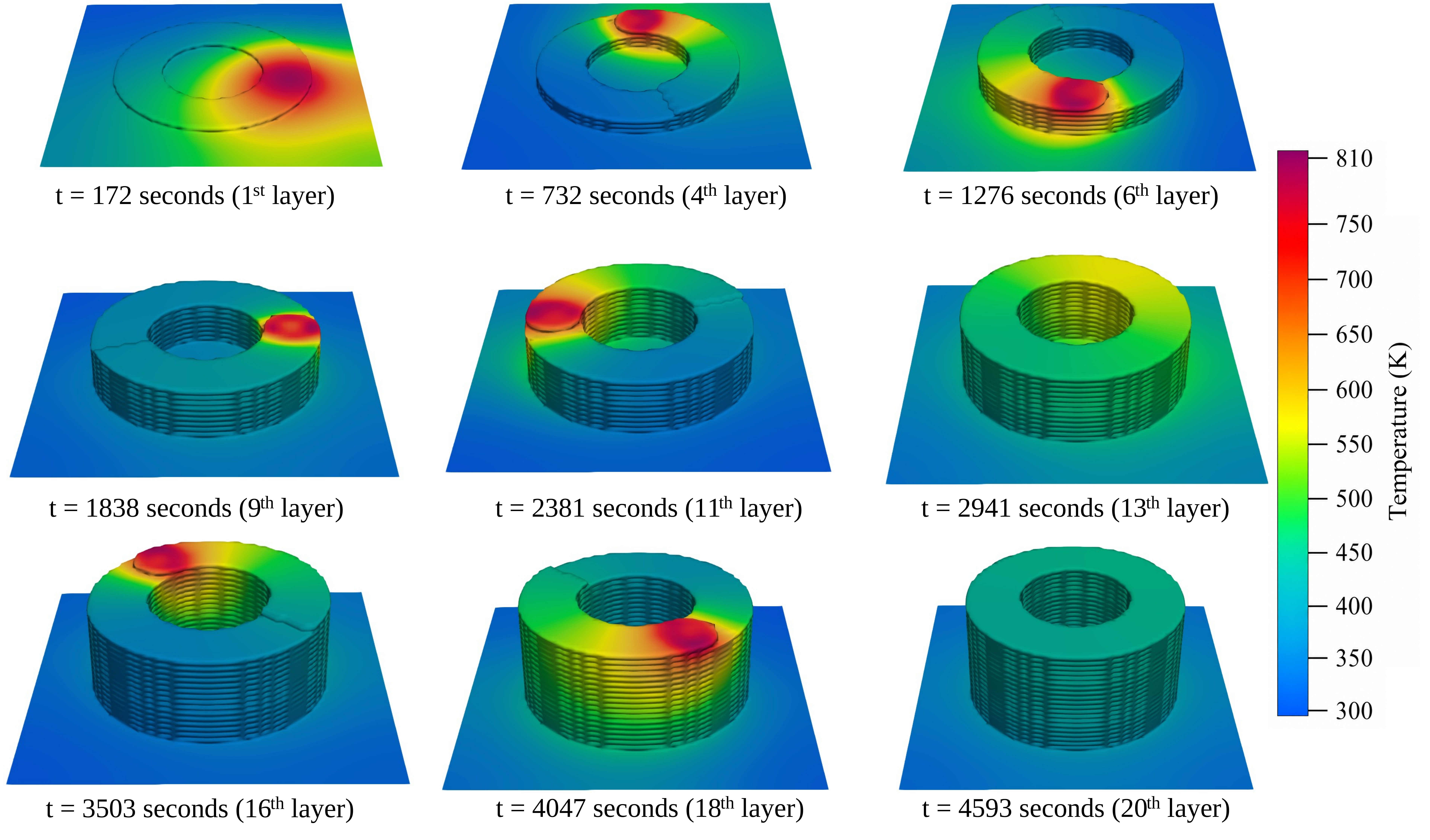}
    \caption{Transient temperature evolution during the 20-layer hollow cylindrical AFSD build at selected stages of deposition.}
    \label{fig:cylinder_temperature}
\end{figure}

Figure~\ref{fig:cylinder_temperature} presents selected temperature fields throughout the 20-layer deposition sequence. The localized high-temperature region follows the circumferential tool path and produces a thermal wake behind the instantaneous deposition location. During the initial layers, the substrate acts as the primary heat sink, resulting in rapid heat transfer away from the deposited material. As the build height increases, heat retained within the previously deposited layers contributes increasingly to the thermal response of subsequent passes.

Although repeated deposition produces a circumferentially continuous heat-affected region, the maximum temperature remains concentrated near the instantaneous tool position rather than becoming uniform around the wall. The thermal response therefore retains the spatial asymmetry associated with the moving heat source throughout the complete build. The results also demonstrate that the fixed-grid geometry representation captures progressive deposition along a closed curved path without layer-wise remeshing.

\subsubsection{Influence of Tool Rotational Speed on the Layer~1 Thermal History}
\label{subsec:layer1_rpm}
\begin{figure}[h!]
    \centering
    \includegraphics[width=1\textwidth]{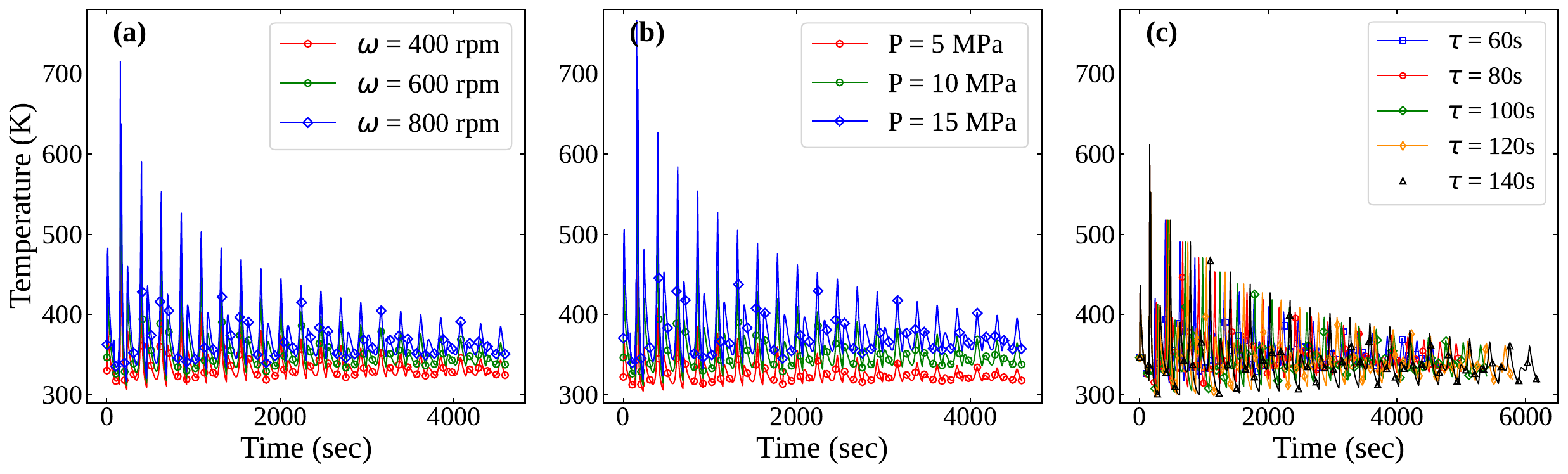}
    \caption{Layer~1 temperature histories for the 20-layer hollow cylindrical AFSD build under three parametric sweeps: (a) tool rotational speed (400, 600, and 800~rpm), (b) applied pressure (5--15~MPa) at 600~rpm and a 60~s dwell, and (c) inter-layer dwell time (60--140~s) at 600~rpm and 10~MPa.}
    \label{fig:layer1_rpm_temperature}
\end{figure}

Figure~\ref{fig:layer1_rpm_temperature}(c) compares the Layer~1 temperature histories obtained at rotational speeds of 400, 600, and 800~rpm. Each tool passage above the circumferential probe location produces a distinct reheating peak, with higher rotational speeds consistently generating higher temperatures. The 800~rpm condition produces the largest peak temperatures, followed by the 600 and 400~rpm conditions.
The magnitude of the reheating peaks decreases as additional layers are deposited because the vertical distance between the active deposition layer and the Layer~1 probe progressively increases. Consequently, the sharp temperature excursions observed during the early layers become increasingly attenuated. A gradual temperature increase occurs near the end of the build for all three rotational speeds. Unlike the earlier tool-passage peaks, this increase is smooth and is attributed to delayed heat redistribution from the upper deposited layers toward the lower portion of the cylinder.
% NOTE: The pressure and dwell-time cases were generated using the local
% implementation during ARC maintenance and should be reverified on ARC
% before submission.
\subsubsection{Influence of Applied Pressure on the Layer~1 Thermal History}
\label{subsec:cylinder_pressure}
Figure~\ref{fig:layer1_rpm_temperature}(b) presents the Layer~1 temperature histories obtained at applied pressures of 5, 10, and 15~MPa while maintaining a rotational speed of 600~rpm and a dwell time of 60~s. Increasing the applied pressure raises the temperature attained during each reheating event but does not alter the timing of the peaks. This response is consistent with the dependence of the prescribed frictional heat input on the applied pressure. The temperature ordering remains consistent throughout the deposition sequence, with the 15~MPa condition producing the highest temperatures and the 5~MPa condition producing the lowest.
\subsubsection{Influence of Inter-Layer Dwell Time on the Layer~1 Thermal History}
\label{subsec:cylinder_dwell}
Figure~\ref{fig:layer1_rpm_temperature}(a) compares the Layer~1 temperature histories for inter-layer dwell times ranging from 60 to 140~s at a rotational speed of 600~rpm and an applied pressure of 10~MPa. Increasing the dwell time progressively delays each subsequent reheating event, causing the temporal separation between the cases to accumulate over the 20-layer deposition sequence.
In contrast to its pronounced effect on the timing of the thermal cycles, dwell time produces a comparatively modest change in the corresponding peak temperatures. The principal influence of the dwell period is therefore the additional cooling interval between successive layers and the associated extension of the overall build duration.

\subsubsection{Final residual stress Distribution}

\begin{figure}[h!]
\centering
\includegraphics[width=0.9\linewidth]{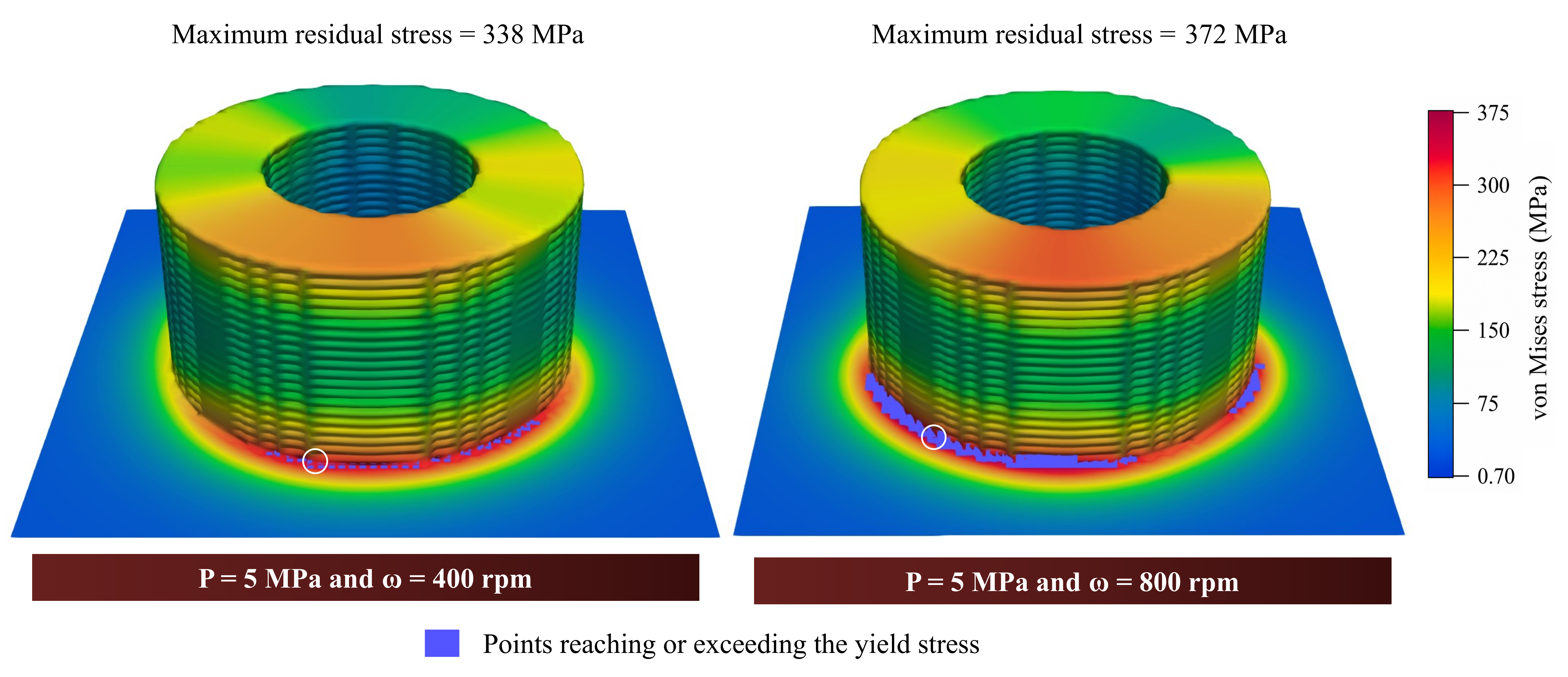}
\caption{Final von Mises residual stress distributions predicted for the 20-layer hollow cylindrical AFSD build at an applied pressure of 5~MPa and tool rotational speeds of 400 and 800~rpm. Points reaching or exceeding the yield stress are colored purple. The maximum stress points are indicated with a white circle.}
\label{fig:cylinder_residual_stress}
\end{figure}

The final von Mises residual stress fields are shown in Figure~\ref{fig:cylinder_residual_stress} for tool rotational speeds of 400 and 800~rpm at an applied pressure of 5~MPa. For both processing conditions, the residual stress distribution is strongly nonuniform, with the largest stresses concentrated around the wall--substrate interface near the base of the cylindrical deposit. A pronounced circumferential high-stress band develops in this region as the deposited wall contracts against the mechanical constraint imposed by the substrate. The stress magnitude decreases progressively away from the wall--substrate junction, producing a radial gradient across the substrate, while regions farther from the deposited wall remain at comparatively low stress levels. Through the cylindrical wall, alternating circumferential stress bands and localized azimuthal variations are observed, reflecting the layer-wise deposition history and the spatially varying thermal-severity field used to scale the RVE-calibrated inherent strain. The 800~rpm condition produces a broader high-stress region near the wall--substrate interface than the 400~rpm condition, consistent with the greater thermal input associated with the higher rotational speed. The purple regions in Figure~\ref{fig:cylinder_residual_stress} identify locations where the predicted von Mises residual stress reaches or exceeds the yield-stress threshold and are concentrated primarily near the lower outer circumference of the deposited wall and the wall--substrate junction. The larger extent of this region at 800~rpm indicates that increasing rotational speed not only increases the peak residual stress but also expands the portion of the structure subjected to stresses near or above yielding. The white circles identify the locations of the maximum predicted residual stress, which increases from 338~MPa at 400~rpm to 372~MPa at 800~rpm.

\begin{figure}[htbp]
\centering
\includegraphics[width=\textwidth]{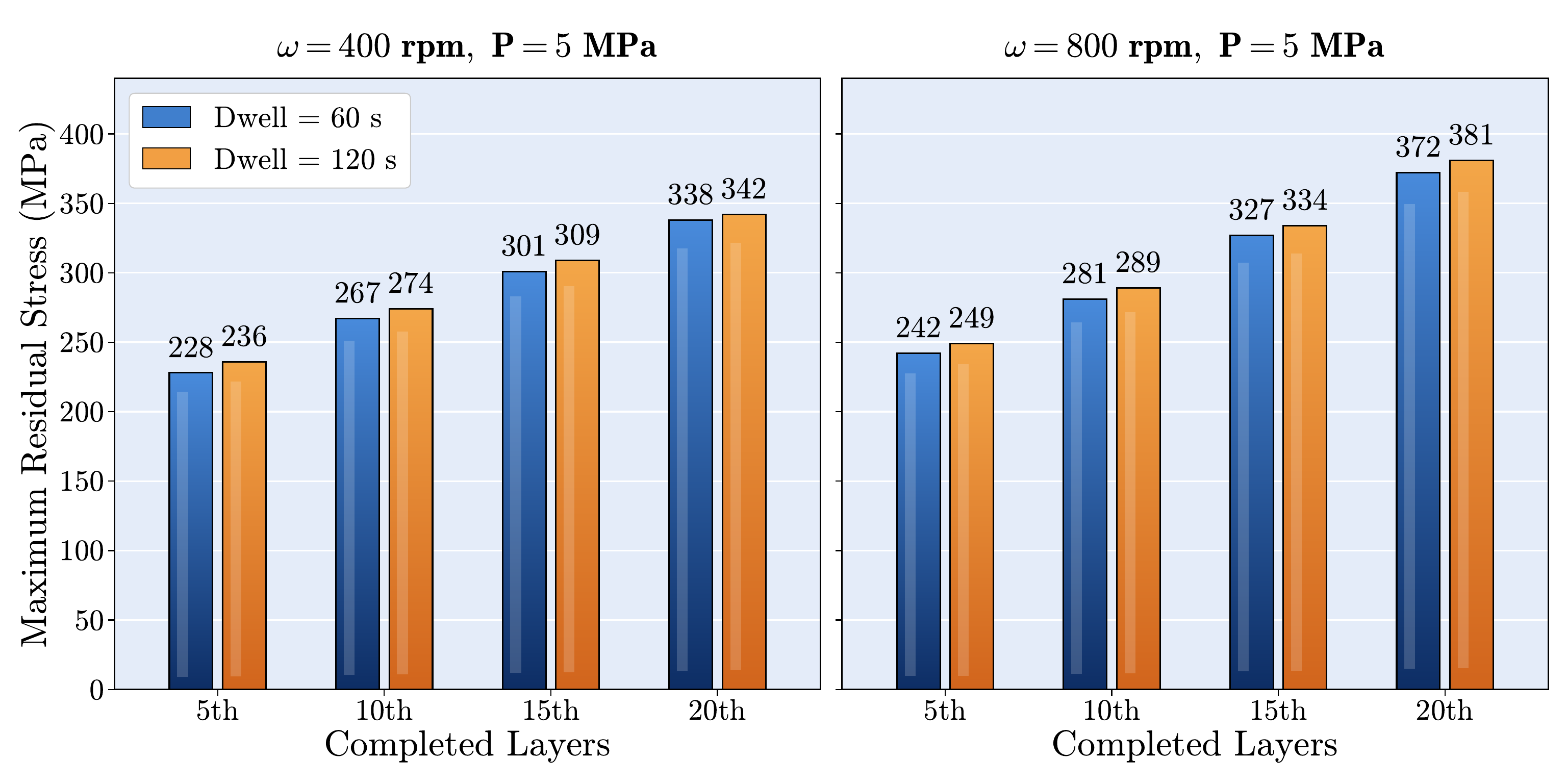}
\caption{Comparison of maximum von Mises residual stress after completion of the 5th, 10th, 15th, and 20th layers of the hollow cylindrical AFSD build for inter-layer dwell times of 60~s and 120~s, at tool rotational speeds of (left) 400~rpm and (right) 800~rpm, with an applied pressure of 5~MPa.}
\label{fig:cylinder_dwell_residual_stress}
\end{figure}

The final residual stress magnitude reflects the stress state only at the end of the complete 20-layer build. To assess how the residual stress accumulates as deposition progresses, and how this accumulation is affected by the inter-layer dwell time, the maximum von Mises residual stress predicted by the ROM/ISM framework is tracked after the 5th, 10th, 15th, and 20th completed layers for two dwell times (60~s and 120~s), at rotational speeds of 400 and 800~rpm and a fixed applied pressure of 5~MPa. The results are shown in Figure~\ref{fig:cylinder_dwell_residual_stress}. For both rotational speeds, the maximum residual stress increases monotonically with the number of completed layers, growing from approximately 228--249~MPa after five layers to 338--381~MPa after the full 20-layer build. This progressive accumulation reflects the increasing mechanical constraint imposed by the growing deposited wall and substrate, together with the cumulative contribution of the thermally scaled eigenstrain field from successive deposition passes.
At every layer count, the 800~rpm condition produces a higher maximum residual stress than the 400~rpm condition, consistent with the greater frictional heat input and correspondingly larger thermal-severity.
Increasing the dwell time from 60~s to 120~s raises the maximum residual stress at every layer count and rotational speed considered, with the 20-layer build reaching 342~MPa (400~rpm) and 381~MPa (800~rpm) at the longer dwell, compared with 338~MPa and 372~MPa at the shorter dwell. The absolute increase associated with the longer dwell is modest and fairly consistent across layer counts and rotational speeds, ranging from approximately 4 to 9~MPa, corresponding to a relative increase of roughly 2--3.5\% that shows a mild tendency to decrease at higher layer counts as the baseline stress level grows. This behavior is qualitatively consistent with the trend reported by Ruvo et al.~\cite{ruvo2026surface} for the AA2024 thin-wall build, where extended inter-layer dwell time was shown to increase the accumulated residual stress by allowing more complete cooling, and therefore greater thermal contraction, between successive depositions, although the effect observed here for the cylindrical build is smaller in magnitude. The present results indicate that this dwell-time sensitivity persists for a geometrically more complex, multi-layer cylindrical build and is superimposed on, rather than coupled to, the layer-count and rotational-speed dependence of the residual stress.

\section{Limitations and Future Directions}
\label{sec:limitations}

The principal limitations of the proposed framework and the corresponding directions for future work are summarized as follows:

\begin{itemize}

    \item \textbf{Inherent strain and elastic mechanical approximation:} The model applies a thermally scaled inherent strain followed by a single elastic equilibrium solve and therefore neglects the incremental thermo-elasto-plastic response and direct tool-induced pressure and contact loading. Because the part-scale mechanical analysis is purely elastic, the predicted stresses are not redistributed through plastic deformation when the von Mises stress exceeds the material yield-stress threshold. Consequently, regions exceeding this threshold should be interpreted as elastic overstress indicators rather than direct predictions of plastically yielded material. This limitation is particularly relevant to the cylindrical build, for which the predicted maximum residual stresses reach 338~MPa and 372~MPa at 400 and 800~rpm, respectively. These approximations, together with RVE averaging, likely contribute to differences between the predicted and physically realizable residual stress fields. Future work should incorporate an elasto-plastic part-scale mechanical correction, pressure-dependent RVE calibration, or a locally coupled process-zone model.

    \item \textbf{B-spline resolution near embedded boundaries:} The quadratic B-spline approximation is defined on a Cartesian background domain that extends beyond the physical material boundary, since the basis functions span neighboring knot intervals. Near embedded boundaries, the finite basis resolution and placement of the outer knot spans can limit the representation of highly localized temperature gradients and may contribute to slight smoothing or underprediction of local peak temperatures. Future work will investigate adaptive B-spline refinement and improved treatment of the background basis near the moving heat source and material boundaries.

    \item \textbf{Simplified heat-source model:} The frictional heat input is prescribed using process parameters and simplified interface assumptions. Variations in sticking and sliding, tool--workpiece contact, material flow, and deformation-generated heat are not explicitly resolved. Future work should incorporate experimentally calibrated heat-partition parameters or coupling with a local process-zone model.

    \item \textbf{Future developments:} Planned extensions include adaptive B-spline refinement near the moving heat source, automatic activation from tool-path data, uncertainty quantification, experimental calibration of the thermal-severity mapping, and process-parameter optimization. Further plan is to explore closed-loop control of the AFSD process using a large language model (LLM) agent framework, following previous related work on LLM agent-guided stochastic control of LPBF~\cite{sarker2026llmafsd}.

\end{itemize}

\section{ Conclusions}
\label{sec:conclusions}

A data-free and geometry-aware reduced-order framework is developed for part-scale thermal and residual stress simulation of AFSD. The transient temperature field is represented using tensor-product quadratic B-splines on a fixed Cartesian grid, while an RVE-calibrated inherent strain model is used to predict the final residual stress field. The thermal model reproduces the principal heating, peak-temperature, cooling, and inter-layer heat-accumulation trends observed in published numerical and experimental results. For the four-layer thin-wall build, the thermal solution requires approximately 21.6~s compared with 806~s for the finite element model, while the complete thermomechanical framework requires approximately 86.4~s compared with 2406~s, corresponding to speed-ups of approximately 37.3 and 27.8 times, respectively. The predicted maximum residual stresses are 231 and 242~MPa at rotational speeds of 400 and 800~rpm, respectively, compared with finite element values of 247 and 259~MPa. For the 20-layer hollow cylindrical build, the reduced-order thermal simulations requires approximately 0.148--0.152~h compared with 1.800~h for the finite element model, corresponding to thermal speed-ups of approximately 11.8--12.2 times. When the mechanical analysis is included, the complete ROM/ISM framework requires approximately 0.346--0.355~h compared with 5.400~h for the full finite element thermomechanical analysis, providing overall speed-ups of approximately 15.2--15.6 times. The final cylindrical residual stress field is strongly nonuniform, with a maximum von Mises stress of approximately 372~MPa near the lower outer circumference of the deposited wall and the wall--substrate junction.

Overall, the proposed framework substantially reduces both thermal and mechanical computational costs while retaining the dominant part-scale temperature and residual stress behavior. The demonstrated performance for both the four-layer thin-wall benchmark and the 20-layer hollow cylindrical build indicates that the framework remains computationally efficient as the deposition history becomes longer and the geometry becomes more complex. It therefore provides an efficient physics-based tool for process-parameter studies, large-scale data generation, and rapid assessment of geometrically evolving AFSD builds.

\section*{Acknowledgements}
S. Saha gratefully acknowledges start-up funding by the Kevin T. Crofton Department of Aerospace and Ocean Engineering, Virginia Tech, and the Professor Ella M. Atkins Faculty Fellowship. The authors acknowledge Advanced Research Computing at Virginia Tech for providing computational resources and technical support that contributed to the results reported in this paper.

\section*{Declaration of generative AI and AI-assisted technologies in the manuscript preparation process}

During the preparation of this work, the authors used ChatGPT and Claude to improve readability and language. After using these tools, the authors reviewed and edited the content as needed and take full responsibility for the content of the publication.

\clearpage
\appendix

\section{Nomenclature}
\label{app:nomenclature}

\begin{table}[h!]
\centering
\caption{Symbols used in the mathematical formulation.}
\label{tab:nomenclature}
\begin{tabular}{ll}
\toprule
Symbol & Definition \\
\midrule
$B_i$ & One-dimensional B-spline basis function \\
$C_p$ & Specific heat capacity \\
$\mathbf{C}$ & Convective heat-loss matrix \\
$\mathbb{C}$ & Elastic stiffness tensor \\
$E$ & Young's modulus \\
$\mathbf{f}$ & Heat-input vector \\
$h$ & Convective heat-transfer coefficient \\
$H_{\epsilon}$ & Smoothed Heaviside function \\
$k$ & Thermal conductivity \\
$\mathbf{K}$ & Thermal conduction matrix \\
$\mathbf{K}_u$ & Mechanical stiffness matrix \\
$\mathbf{M}$ & Thermal capacity matrix \\
$N_A$ & Tensor-product B-spline basis function \\
$q$ & Volumetric heat input \\
$s$ & inherent strain scaling factor \\
$T$ & Temperature \\
$T_{\max}$ & Maximum local temperature \\
$T_{\mathrm{ref}}$ & Reference temperature \\
$T_{\infty}$ & Ambient or far-field temperature \\
$\mathbf{u}$ & Displacement field \\
$v$ & Tool traverse speed \\
$w$ & Test function \\
$\alpha$ & Coefficient of thermal expansion \\
$\gamma$ & Diffuse surface measure \\
$\delta_{\epsilon}$ & Regularized Dirac delta function \\
$\epsilon_{\mathrm{rad}}$ & Surface emissivity \\
$\mathbf{n}$ & Outward unit normal \\
$\boldsymbol{\varepsilon}$ & Total strain tensor \\
$\boldsymbol{\varepsilon}^{*}$ & Imposed inherent strain tensor \\
$\eta$ & Smoothed material indicator \\
$\boldsymbol{\theta}$ & B-spline temperature coefficients \\
$\nu$ & Poisson's ratio \\
$\rho$ & Density \\
$\boldsymbol{\sigma}$ & Stress tensor \\
$\sigma_{\mathrm{SB}}$ & Stefan--Boltzmann constant \\
$\sigma_y$ & Yield stress \\
$\phi$ & Signed distance function \\
$\chi$ & Thermal severity index \\
$\chi_0$ & Reference thermal severity at RVE calibration condition \\
$T_{\chi,\max}$ & Temperature cap for thermal-severity evaluation \\
$\Omega$ & Cartesian background domain \\
$\Omega_m$ & Material domain \\
$\Gamma$ & Material boundary \\
\bottomrule
\end{tabular}
\end{table}

\section{Mathematical Formulation}
\label{app:mathematical_formulation}

\subsection{Handling CAD Geometry}

To handle arbitrary CAD geometry without generating a body-fitted finite element mesh, the part is embedded into a fixed Cartesian background domain using a signed distance function (SDF). In this approach, the CAD surface is first converted into a scalar distance field $\phi(\mathbf{x})$ defined on the background grid. This background grid is simply a rectangular computational grid that encloses the entire part. The grid does not follow the CAD boundary. Instead, the signed distance function determines which grid points are inside the material, outside the material, or near the CAD surface. The sign of $\phi$ identifies whether a point is inside, on, or outside the material domain:
\begin{align}
    \phi(\mathbf{x}) < 0, &\quad \mathbf{x} \in \Omega_m, \\
    \phi(\mathbf{x}) = 0, &\quad \mathbf{x} \in \Gamma, \\
    \phi(\mathbf{x}) > 0, &\quad \mathbf{x} \notin \Omega_m,
\end{align}
where $\Omega_m$ is the material domain and $\Gamma$ is the CAD boundary.

This representation is useful for complex or changing geometries because the computational grid remains Cartesian while the CAD shape enters the governing equations only through $\phi(\mathbf{x})$. Therefore, volume integrals are restricted to the material region by a smoothed material indicator,
\begin{equation}
    \eta(\mathbf{x}) = H_{\epsilon}(-\phi(\mathbf{x})), \qquad
    H_{\epsilon}(-\phi)=
    \begin{cases}
    1, & \phi \le -\epsilon,\\
    \dfrac{1}{2}\left(1-\dfrac{\phi}{\epsilon}
    -\dfrac{1}{\pi}\sin\left(\dfrac{\pi\phi}{\epsilon}\right)\right),
    & |\phi|<\epsilon,\\
    0, & \phi \ge \epsilon.
    \end{cases}
\end{equation}
Here, $\eta\approx 1$ inside the CAD part, $\eta\approx 0$ outside the part, and the transition is smoothed over a narrow band of thickness controlled by $\epsilon$.

Similarly, heat-loss boundary terms are applied on the CAD surface using the diffuse surface measure
\begin{equation}
    \gamma(\mathbf{x}) =
    \delta_{\epsilon}(\phi(\mathbf{x}))|\nabla\phi(\mathbf{x})|, \qquad
    \delta_{\epsilon}(\phi)=
    \begin{cases}
    \dfrac{1}{2\epsilon}
    \left[
    1+\cos\left(\dfrac{\pi\phi}{\epsilon}\right)
    \right],
    & |\phi|<\epsilon,\\
    0, & |\phi|\ge\epsilon.
    \end{cases}
\end{equation}
The factor $|\nabla \phi|$ converts the narrow-band volume integral into an equivalent surface measure. As a result, both material-volume terms and CAD-surface heat-loss terms can be evaluated on the same background grid, avoiding repeated meshing or remeshing for complex CAD parts.

\begin{figure}[h!]
\centering
\includegraphics[width=1\linewidth]{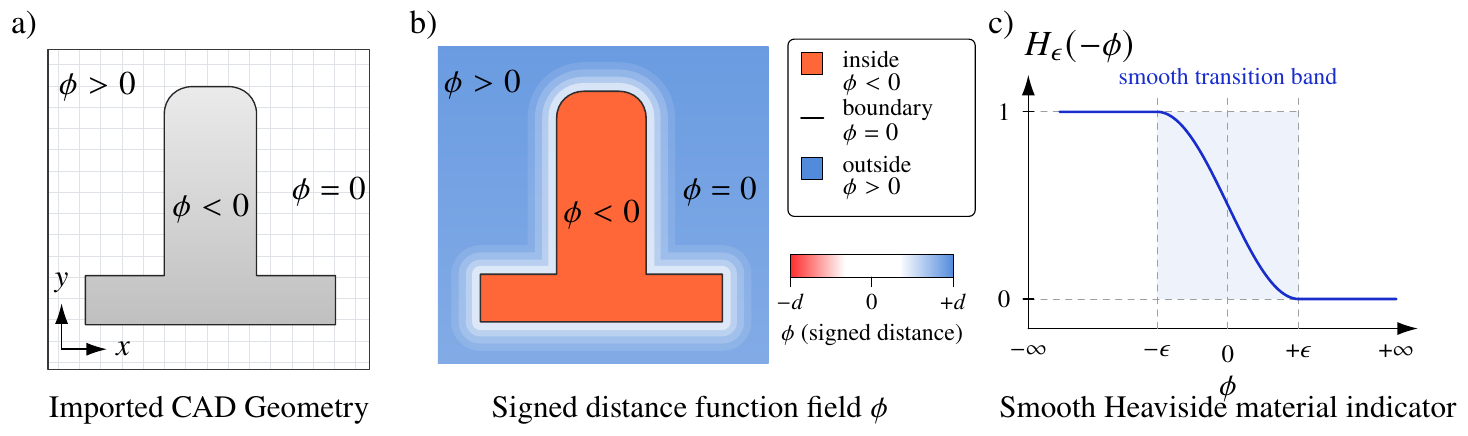}
\caption{Signed-distance-function-based representation of the part geometry on a fixed Cartesian background grid: (a) imported CAD geometry and the sign convention for the distance field, (b) signed distance function $\phi$, with negative values inside the material and positive values outside, and (c) smooth Heaviside material indicator used to transition continuously across the embedded material boundary.}
\label{fig:sdf_embedding}
\end{figure}

\subsection{Thermal Solver with Separable Approximation}

The transient temperature field during AFSD is governed by the heat conduction equation
\begin{equation}
    \rho C_p(T)\frac{\partial T}{\partial t}
    =
    \nabla \cdot \left(k(T)\nabla T\right)
    + q(\mathbf{x},t),
\end{equation}
where $\rho$ is density, $C_p(T)$ is the temperature-dependent specific heat,
$k(T)$ is the temperature-dependent thermal conductivity, and
$q(\mathbf{x},t)$ is the AFSD heat input. Heat loss from the free surface is represented by convection and radiation,
\begin{equation}
    -k\nabla T \cdot \mathbf{n}
    =
    h(T-T_{\infty})
    +
    \epsilon_{\mathrm{rad}}\sigma_{\mathrm{SB}}
    \left(T^4-T_{\infty}^4\right),
\end{equation}
where $h$ is the convection coefficient, $T_{\infty}$ is the ambient temperature,
$\epsilon_{\mathrm{rad}}$ is the surface emissivity, and $\sigma_{\mathrm{SB}}$ is the Stefan--Boltzmann constant.

Using the embedded CAD representation, the governing equation is written over the Cartesian background domain $\Omega$ rather than over a body-fitted mesh. The material volume terms are weighted by the smoothed indicator $\eta(\mathbf{x})$, while the surface heat-loss terms are weighted by the diffuse surface measure $\gamma(\mathbf{x})$. For a test function $w$, the weak form becomes
\begin{align}
    \int_{\Omega} \rho C_p(T)\eta w \frac{\partial T}{\partial t} \, dV
    &+
    \int_{\Omega} k(T)\eta \nabla w \cdot \nabla T \, dV  \nonumber\\
    &+
    \int_{\Omega} h\gamma w(T-T_{\infty}) \, dV
    +
    \int_{\Omega} \epsilon_{\mathrm{rad}}\sigma_{\mathrm{SB}}\gamma w
    \left(T^4-T_{\infty}^4\right) \, dV
    =
    \int_{\Omega} \eta w q \, dV .
\end{align}

The reduced thermal field is approximated using a tensor-product quadratic B-spline basis,
\begin{equation}
    T_h(x,y,z,t)
    =
    \sum_{i=1}^{n_x}
    \sum_{j=1}^{n_y}
    \sum_{k=1}^{n_z}
    \theta_{ijk}(t)
    B_i(x)B_j(y)B_k(z),
\end{equation}
where $\theta_{ijk}(t)$ are the unknown temperature coefficients. The approximation is separable in space because each three-dimensional basis function is constructed as a product of one-dimensional basis functions,
\begin{equation}
    N_{ijk}(x,y,z)=B_i(x)B_j(y)B_k(z).
\end{equation}
Its gradient is therefore evaluated as
\begin{equation}
    \nabla N_{ijk}
    =
    \begin{bmatrix}
    B_i'(x)B_j(y)B_k(z)\\
    B_i(x)B_j'(y)B_k(z)\\
    B_i(x)B_j(y)B_k'(z)
    \end{bmatrix}.
\end{equation}

Choosing the test functions from the same B-spline space gives the Galerkin semi-discrete system
\begin{equation}
    \mathbf{M}(\boldsymbol{\theta})\dot{\boldsymbol{\theta}}
    +
    \mathbf{K}(\boldsymbol{\theta})\boldsymbol{\theta}
    +
    \mathbf{C}\boldsymbol{\theta}
    +
    \mathbf{r}_{\mathrm{rad}}(\boldsymbol{\theta})
    =
    \mathbf{f}(t),
\end{equation}
where
\begin{align}
    M_{AC} &= \int_{\Omega} \rho C_p(T)\eta N_A N_C \, dV, \\
    K_{AC} &= \int_{\Omega} k(T)\eta \nabla N_A \cdot \nabla N_C \, dV, \\
    C_{AC} &= \int_{\Omega} h \gamma N_A N_C \, dV.
\end{align}
Here, $\mathbf{M}$ is the thermal capacity matrix, $\mathbf{K}$ is the conduction matrix, $\mathbf{C}$ is the convective heat-loss matrix, $\mathbf{r}_{\mathrm{rad}}$ is the nonlinear radiation contribution, and $\mathbf{f}$ is the heat-input vector.

All integrals are evaluated on the Cartesian quadrature grid. For example, a generic volume integral is approximated as
\begin{equation}
    \int_{\Omega} g(\mathbf{x})\,dV
    \approx
    \sum_{m=1}^{N_q} g(\mathbf{x}_m)w_m,
\end{equation}
where $\mathbf{x}_m$ and $w_m$ are the quadrature point and weight. The embedded weights $\eta_m$ and $\gamma_m$ determine whether each quadrature point contributes as material volume, surface boundary, or exterior region.

Using backward Euler time integration, the linearized system at time step $n+1$ is written as
\begin{equation}
    \left(
    \frac{\mathbf{M}}{\Delta t}
    +
    \mathbf{K}
    +
    \mathbf{C}
    \right)
    \boldsymbol{\theta}^{n+1}
    =
    \mathbf{f}^{n+1}
    +
    \frac{\mathbf{M}}{\Delta t}
    \boldsymbol{\theta}^{n}
    -
    \mathbf{r}_{\mathrm{rad}}^{n+1},
\end{equation}
where the radiation term may be evaluated explicitly, lagged from the previous iterate, or included through a nonlinear update. In the present implementation, the system is solved in matrix-free form: the action of the operator on a vector is computed directly without assembling the full global matrix. This reduces memory cost and enables efficient GPU acceleration.

\subsection{Modified Inherent Strain Method}
\label{subsec:rve_calibration}

The thermal solution is mapped to the mechanical voxel grid by sampling the B-spline temperature field at element centroids. For each mechanical voxel, the peak temperature is computed as
\begin{equation}
    T_{\max}(\mathbf{x}) = \max_t T_h(\mathbf{x},t).
\end{equation}

A dimensionless thermal severity index is then defined as
\begin{equation}
    \chi(\mathbf{x})
    =
    \frac{
    E(T)\alpha(T)|T_{\max}(\mathbf{x})-T_{\mathrm{ref}}|
    }
    {
    (1-\nu)\sigma_y(T)
    },
\end{equation}
where $E$ is Young's modulus, $\alpha$ is the coefficient of thermal expansion, $\nu$ is Poisson's ratio, and $\sigma_y$ is the temperature-dependent yield stress. The reference temperature is taken equal to the ambient temperature, $T_{\mathrm{ref}}=T_{\mathrm{amb}}=T_\infty=298$~K, consistent with the initial and far-field conditions of Eqs.~\eqref{eq:thinwall_initial_temperature} and \eqref{eq:main_thermal_boundary}. This index compares a thermal stress scale with the local yield strength. Regions with larger thermal severity are therefore assigned larger inherent strain.

To avoid unphysical amplification at very high temperatures, the temperature used in evaluating $\chi$ is capped as
\begin{equation}
    T_{\mathrm{cap}}
    =
    \min\left(\max(T_{\max},T_{\mathrm{ref}}),T_{\chi,\max}\right).
\end{equation}

The local inherent strain field is obtained by scaling a representative RVE-calibrated inherent strain,
\begin{equation}
    \boldsymbol{\varepsilon}^{*}(\mathbf{x})
    =
    s(\mathbf{x})\boldsymbol{\varepsilon}^{*}_{\mathrm{avg}},
\end{equation}
where
\begin{equation}
    s(\mathbf{x})
    =
    \frac{\max(\chi(\mathbf{x})-1,0)}{\chi_0-1}.
\end{equation}
Here, $\boldsymbol{\varepsilon}^{*}_{\mathrm{avg}}$ and $\chi_0$ are obtained from the representative volume element calibration. The use of $\max(\chi-1,0)$ ensures that inherent strain is assigned only to regions where the thermal loading exceeds a yield-like threshold.

\subsection{Eigenstrain Solver}

The residual stress problem is formulated using small-strain elasticity with imposed eigenstrain. The strain tensor is
\begin{equation}
    \boldsymbol{\varepsilon}(\mathbf{u})
    =
    \frac{1}{2}
    \left(
    \nabla \mathbf{u}
    +
    \nabla \mathbf{u}^{T}
    \right),
\end{equation}
where $\mathbf{u}$ is the displacement field. The constitutive equation is
\begin{equation}
    \boldsymbol{\sigma}
    =
    \mathbb{C}:
    \left(
    \boldsymbol{\varepsilon}
    -
    \boldsymbol{\varepsilon}^{*}
    \right),
\end{equation}
where $\mathbb{C}$ is the elastic stiffness tensor and $\boldsymbol{\varepsilon}^{*}$ is the imposed inherent strain.

The static equilibrium equation is
\begin{equation}
    \nabla \cdot \boldsymbol{\sigma} = \mathbf{0}.
\end{equation}
The weak form is obtained by multiplying by a virtual displacement field and integrating over the mechanical domain:
\begin{equation}
    \int_{\Omega_m}
    \delta \boldsymbol{\varepsilon}^{T}
    \boldsymbol{\sigma}
    \, dV
    =
    0.
\end{equation}
Substituting the eigenstrain constitutive relation gives
\begin{equation}
    \int_{\Omega_m}
    \delta \boldsymbol{\varepsilon}^{T}
    \mathbb{C}
    \boldsymbol{\varepsilon}
    \, dV
    =
    \int_{\Omega_m}
    \delta \boldsymbol{\varepsilon}^{T}
    \mathbb{C}
    \boldsymbol{\varepsilon}^{*}
    \, dV.
\end{equation}

After finite element discretization using Hex8 voxel elements,
\begin{equation}
    \mathbf{K}_u\mathbf{d}
    =
    \mathbf{f}^{*},
\end{equation}
where
\begin{align}
    \mathbf{K}_u
    &=
    \int_{\Omega_m}
    \mathbf{B}^{T}\mathbb{C}\mathbf{B}
    \, dV, \\
    \mathbf{f}^{*}
    &=
    \int_{\Omega_m}
    \mathbf{B}^{T}\mathbb{C}
    \boldsymbol{\varepsilon}^{*}
    \, dV.
\end{align}
Thus, the inherent strain appears as an equivalent mechanical load. Once the displacement field is solved, the residual stress is recovered from
\begin{equation}
    \boldsymbol{\sigma}
    =
    \mathbb{C}:
    \left(
    \mathbf{B}\mathbf{d}
    -
    \boldsymbol{\varepsilon}^{*}
    \right).
\end{equation}

% The von Mises stress is computed as
% \begin{equation}
%     \sigma_{\mathrm{vm}}
%     =
%     \sqrt{
%     \frac{1}{2}
%     \left[
%     (\sigma_{xx}-\sigma_{yy})^2
%     +
%     (\sigma_{yy}-\sigma_{zz})^2
%     +
%     (\sigma_{zz}-\sigma_{xx})^2
%     +
%     6(\tau_{xy}^2+\tau_{yz}^2+\tau_{xz}^2)
%     \right]
%     }.
% \end{equation}

In the present inherent strain ROM, the default part-scale mechanical solve is elastic with imposed eigenstrain. Plastic deformation is not evolved at every part-scale integration point; instead, the local inelastic process history is condensed into the RVE-calibrated inherent strain field.

\section{Convergence Study}
\label{app:convergence}

A convergence study is performed to determine the spatial resolution required by the B-spline Galerkin thermal ROM. The tensor-product basis resolution is varied over the six levels listed in Table~\ref{tab:dof_levels}, while the process parameters, material properties, time-step settings, and boundary conditions are held fixed. The experimentally validated finite element solution of Ruvo et al.~\cite{ruvo2026surface} was used as the reference.

\begin{table}[h!]
\centering
\caption{B-spline resolutions used in the convergence study.}
\label{tab:dof_levels}
\begin{tabular}{ccc}
\toprule
Level & $(n_x,n_y,n_z)$ & DOF \\
\midrule
1 & $(8,5,4)$   & 420 \\
2 & $(12,7,5)$  & 882 \\
3 & $(16,9,6)$  & 1,584 \\
4 & $(24,14,9)$ & 4,576 \\
5 & $(32,18,12)$ & 9,520 \\
6 & $(56,20,14)$ & 20,416 \\
\bottomrule
\end{tabular}
\end{table}

The error is evaluated as the global root-mean-square temperature difference between the ROM and FEM solutions over all active material points at the four layer-end states:

\begin{equation}
    \|e_T\|_{L^2}
    =
    \left[
    \frac{1}{\sum_{\ell=1}^{N_L}|\mathcal{A}_{\ell}|}
    \sum_{\ell=1}^{N_L}
    \sum_{\mathbf{x}_p\in\mathcal{A}_{\ell}}
    \left(
    T_h(\mathbf{x}_p,t_{\ell})
    -
    T^{\mathrm{FEM}}(\mathbf{x}_p,t_{\ell})
    \right)^2
    \right]^{1/2},
    \label{eq:l2_thermal_error}
\end{equation}

where $N_L=4$, $\mathcal{A}_{\ell}$ denotes the set of active material points at the end of layer $\ell$, and $t_{\ell}$ is the corresponding layer-end time.

\begin{figure}[h!]
    \centering
    \includegraphics[width=\linewidth]{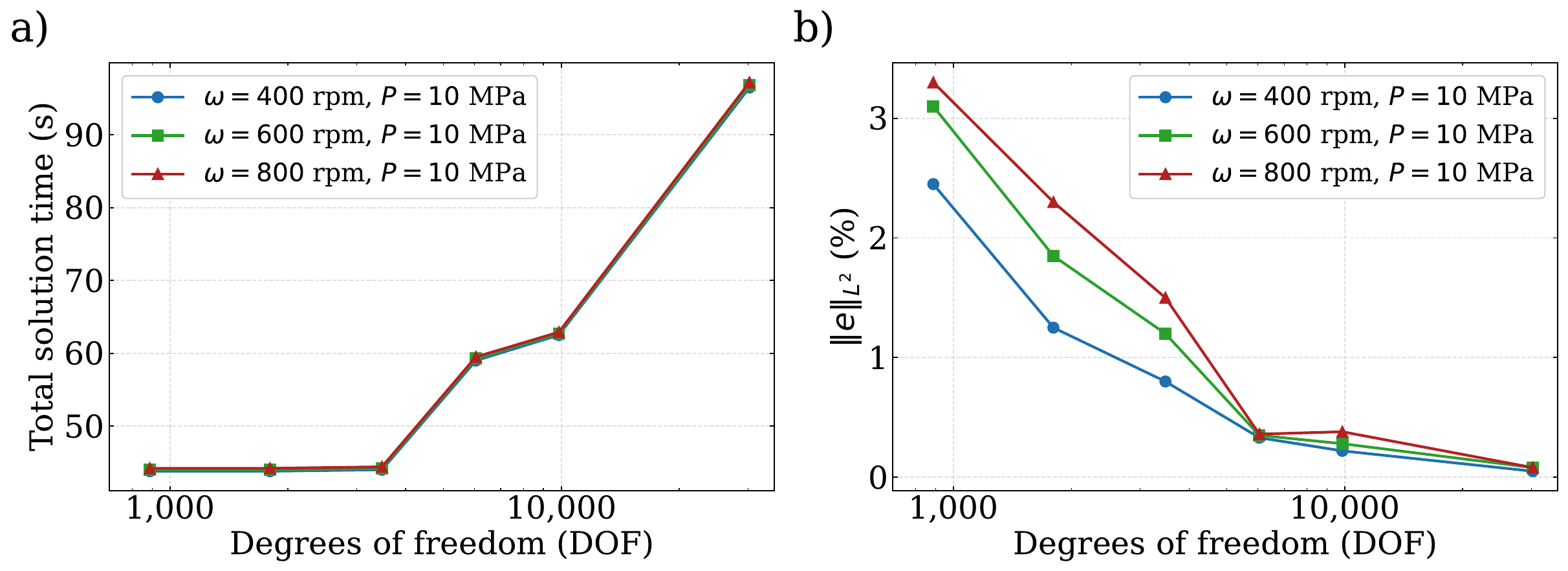}
    \caption{Solution time and thermal error as functions of B-spline degrees of freedom at rotational speeds of 400, 600, and 800~rpm.}
    \label{fig:convergence}
\end{figure}

Figure~\ref{fig:convergence} shows the variation of $\|e_T\|_{L^2}$ with the number of B-spline degrees of freedom for rotational speeds of 400, 600, and 800~rpm. The applied pressure, traverse speed, and dwell time are fixed at 10~MPa, 100~mm/min, and 60~s, respectively. The error decreases rapidly from Level~1 to Level~3 and then approaches an approximately constant value of 26~K. Further refinement to 20,416~DOF produces only marginal improvement, indicating that B-spline spatial discretization is no longer the dominant source of discrepancy. 

\section{Material Properties}
\label{app:material_props}

The AA2024 properties used in the thermal and mechanical simulations are summarized in Tables~\ref{tab:material_constants}--\ref{tab:mechanical_properties}. Temperature-dependent quantities are linearly interpolated between the listed values and held constant beyond the table endpoints. The properties are adopted from the material model used by Ruvo et al.~\cite{ruvo2026surface}.

\begin{table}[h!]
\centering
\caption{Constant properties and phase-transition temperatures of AA2024.}
\label{tab:material_constants}
\begin{tabular}{lc}
\toprule
Quantity & Value \\
\midrule
$\rho$ (kg\,m$^{-3}$) & 2800 \\
$T_{\mathrm{solidus}}$ (K) & 763 \\
$T_{\mathrm{liquidus}}$ (K) & 913 \\
\bottomrule
\end{tabular}
\end{table}

\begin{table}[h!]
\centering
\caption{Temperature-dependent thermal properties of AA2024.}
\label{tab:thermal_properties}
\begin{tabular}{ccc}
\toprule
$T$ (K) &
$k$ (W\,m$^{-1}$\,K$^{-1}$) &
$C_p$ (J\,kg$^{-1}$\,K$^{-1}$) \\
\midrule
293 & 164 & 881 \\
373 & 182 & 927 \\
473 & 194 & 1047 \\
573 & 202 & 1130 \\
673 & 210 & 1210 \\
773 & 220 & 1300 \\
\bottomrule
\end{tabular}
\end{table}

\begin{table}[h!]
\centering
\caption{Temperature-dependent mechanical properties of AA2024.}
\label{tab:mechanical_properties}
\begin{tabular}{cccc}
\toprule
$T$ (K) &
$E$ (GPa) &
$\sigma_y$ (MPa) &
$\alpha$ ($10^{-5}$ K$^{-1}$) \\
\midrule
298 & 73.0 & 325 & 6.96 \\
373 & 70.0 & 290 & 7.20 \\
423 & 67.0 & 260 & 7.35 \\
473 & 63.0 & 170 & 7.50 \\
523 & 58.0 & 100 & 7.65 \\
573 & --   & 50  & 7.80 \\
623 & --   & 30  & 7.95 \\
673 & --   & 20  & 8.10 \\
723 & --   & --  & 8.25 \\
773 & --   & --  & 8.40 \\
\bottomrule
\end{tabular}
\end{table}


\bibliographystyle{elsarticle-num}
\bibliography{bibliography}



%% For example:

%% \bibitem[Aladro et al.(2015)]{Aladro15} Aladro, R., Martín, S., Riquelme, D., et al. 2015, \aas, 579, A101

%%\end{thebibliography}

\end{document}